\documentclass{article}

\def\shownotes{1}

\usepackage[backend=biber,
            style=numeric-comp,
            bibstyle=authoryear,
            natbib=true,
            maxbibnames=99,
            minalphanames=3,
            maxcitenames=2,
            sorting=ynt,
            uniquename=false,
            uniquelist=false,
            giveninits=true,
            labeldateparts=true,
            sortcites=true,
            backref=true]{biblatex}
\renewcommand{\cite}[1]{\citep{#1}}
\usepackage[hidelinks,breaklinks=true]{hyperref}
\hypersetup{colorlinks=true,linkcolor={DarkRed},citecolor={black},urlcolor={DarkRed}}
\DefineBibliographyStrings{english}{%
  backrefpage = {\hspace{-0.0750cm}},%
  backrefpages = {\hspace{-0.0750cm}},%
}
\DeclareNameAlias{sortname}{family-given}
\makeatletter
\input{numeric.bbx}
\makeatother

\usepackage{macro}
\usepackage[us,12hr]{datetime} %
\usepackage{graphicx, subcaption}
\usepackage{multirow, makecell}
\usepackage[T1]{fontenc}
\usepackage{microtype}
\usepackage{sidecap}
\usepackage{graphics}
\usepackage{graphicx}
\usepackage{booktabs} %
\usepackage{enumitem}
\makeatletter %
\renewcommand\Large{\@setfontsize\Large{16.5pt}{0}}
\makeatother
\makeatletter
\def\thanksnosymbol#1{\protected@xdef\@thanks{\@thanks
        \protect\footnotetext{#1}}}
\makeatother
\usepackage{hyperref}
\usepackage{cleveref}
\usepackage{wrapfig}
\usepackage{breakcites}
\usepackage{algorithm}
\usepackage{algorithmic}

\usepackage{listings}
\usepackage{color}

\definecolor{dkgreen}{rgb}{0,0.6,0}
\definecolor{gray}{rgb}{0.5,0.5,0.5}
\definecolor{mauve}{rgb}{0.58,0,0.82}

\PassOptionsToPackage{hyphens}{url}

\usepackage[final,nonatbib]{neurips_data_2023}
\usepackage{amsfonts}       %
\usepackage{nicefrac}       %
\usepackage{microtype}      %
\makeatletter
\def\thanksnosymbol#1{\protected@xdef\@thanks{\@thanks
        \protect\footnotetext{#1}}}
\makeatother

\graphicspath{ {./images/} }

\title{Graph Neural Networks for Road Safety Modeling:\\Datasets and Evaluations for Accident Analysis}

\author{%
Abhinav Nippani$^{\ddag}$\thanksnosymbol{$^{\ddag}$First two authors contributed equally. Correspondence can be directed to all authors with the emails above.}, Dongyue Li$^{\ddag}$, Haotian Ju, Haris N. Koutsopoulos, Hongyang R. Zhang\\
Northeastern University, Boston\\
\texttt{\{nippani.a, li.dongyu, ju.h, h.koutsopoulos, ho.zhang\}@northeastern.edu}\\
}

\begin{document}

\maketitle
\begin{abstract}
We consider the problem of traffic accident analysis on a road network based on road network connections and traffic volume. Previous works have designed various deep-learning methods using historical records to predict traffic accident occurrences. However, there is a lack of consensus on how accurate existing methods are, and a fundamental issue is the lack of public accident datasets for comprehensive evaluations. This paper constructs a large-scale, unified dataset of traffic accident records from official reports of various states in the US, totaling 9 million records, accompanied by road networks and traffic volume reports. Using this new dataset, we evaluate existing deep-learning methods for predicting the occurrence of accidents on road networks. Our main finding is that graph neural networks such as GraphSAGE can accurately predict the number of accidents on roads with less than 22\% mean absolute error (relative to the actual count) and whether an accident will occur or not with over 87\% AUROC, averaged over states. We achieve these results by using multitask learning to account for cross-state variabilities (e.g., availability of accident labels) and transfer learning to combine traffic volume with accident prediction. Ablation studies highlight the importance of road graph-structural features, amongst other features. Lastly, we discuss the implications of the analysis and develop a package for easily using our new dataset.
\end{abstract}

\section{Introduction}

Graph neural networks are widely used tools for extracting structural relationships from data.
Examples include friendships and interactions on social networks \cite{kipf2016semi,chiang2019cluster}, 3D protein-protein interactions \cite{wu2018moleculenet,morris2020tudataset}, and road connections on traffic networks \cite{li2017diffusion}.
Motivated by the widespread use of graph neural networks, large-scale graph databases and benchmarks have received significant interest in recent studies \cite{hu2020open,hu2021ogb,freitas2020large}.
Existing architecture designs can be abstracted in the mathematical framework of message-passing neural networks \cite{ju2023generalization}.
In practice, the empirical performance of different network designs varies across domains \cite{li2018multi}.
To facilitate the discussion, this paper examines graph neural networks for the important problem of traffic accident risk modeling: Given historical accident records as edge labels on a road network, how well can we predict the number of accident occurrences on roads (e.g., over the next month) using graph-structural and related features?

The importance of modeling traffic accident risks is well-recognized, as many cities propose vision zero plans to eliminate motor vehicle crashes \cite{vz_boston,vz_nyc,vz_lac,vz_bayarea}.
According to CDC \cite{cdc1}, the economic cost of crashes amounted to \$430 billion in 2020.
Developing better modeling tools can identify the underlying risk of accidents at a certain location \cite{he2021inferring}, thus informing policy intervention \cite{balakrishnan2017making}. %

Many studies have analyzed the effect of road features for predicting accident occurrences, such as traffic flow, road network geometry, and rain \cite{persaud1992accident,fridstrom1995measuring,hadi1995estimating}.
For example, a regression analysis has been conducted to quantify the safety effects of the annual average daily traffic (AADT) and rain (among others) using observed crash data in Italy \cite{caliendo2007crash}.
Human, vehicle, and environmental factors have also been incorporated into road accident modeling with autoregressive models \cite{ihueze2018road}.
These studies focus on simple regression models.
The use of deep learning for traffic accident prediction has recently been examined \cite{lin2015novel,ren2018deep,zhou2020riskoracle}.
A heterogeneous convolutional LSTM framework has been developed to predict traffic accidents based on data collected from Iowa state \cite{yuan2018hetero}.
\citet{moosavi2019accident} train a model composed of recurrent and feedforward networks with two million accident record labels across the US.
Despite significant recent interests and strong societal importance, it remains unclear how accurately existing deep learning methods, particularly graph neural networks, can be used to predict accident occurrences on road networks.

A critical challenge in addressing this question is a lack of large-scale traffic accident datasets.
There is a large body of work focusing on traffic forecasting (see, e.g., \citet{li2017diffusion,jiang2022graph,wang2020deep}), but the datasets therein are usually not annotated with accident records.
\citet{moosavi2019accident} construct a repository including over two million accident records collected from map APIs across US states.
By contrast, we construct a new dataset with over nine million traffic accident records spanning eight states, the longest spanning twenty years.
We extract these records from official reports of the Department of Transportation from each state; each comes with the latitude, longitude, and time of day of occurrence.
This is a nontrivial task, as different states publish their data in various formats and APIs, and we unify them into a standard format.
We then collect road networks, road-level AADT, and weather reports aligned with the accident labels in our dataset.

\begin{figure}[t!]
\centering
\begin{minipage}[b]{0.49\textwidth}
 \begin{subfigure}[b]{0.49\textwidth}
     \centering
     \renewcommand\thesubfigure{\alph{subfigure}1}
     \includegraphics[width=\textwidth]{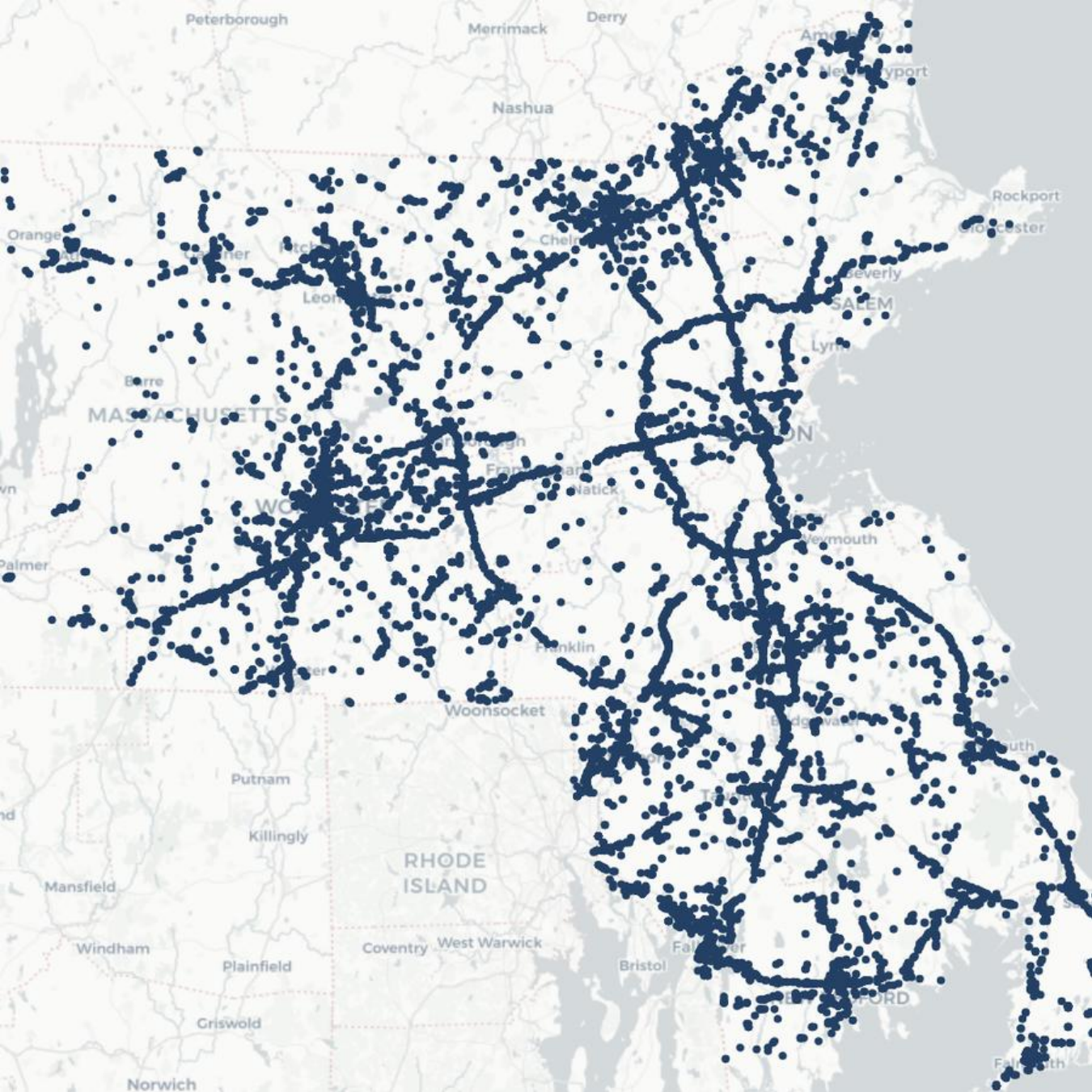}
     \caption{MA, Accidents}
 \end{subfigure}\hspace{0.01in}
 \begin{subfigure}[b]{0.49\textwidth}
     \centering
     \addtocounter{subfigure}{-1}
     \renewcommand\thesubfigure{\alph{subfigure}2}
     \includegraphics[width=\textwidth]{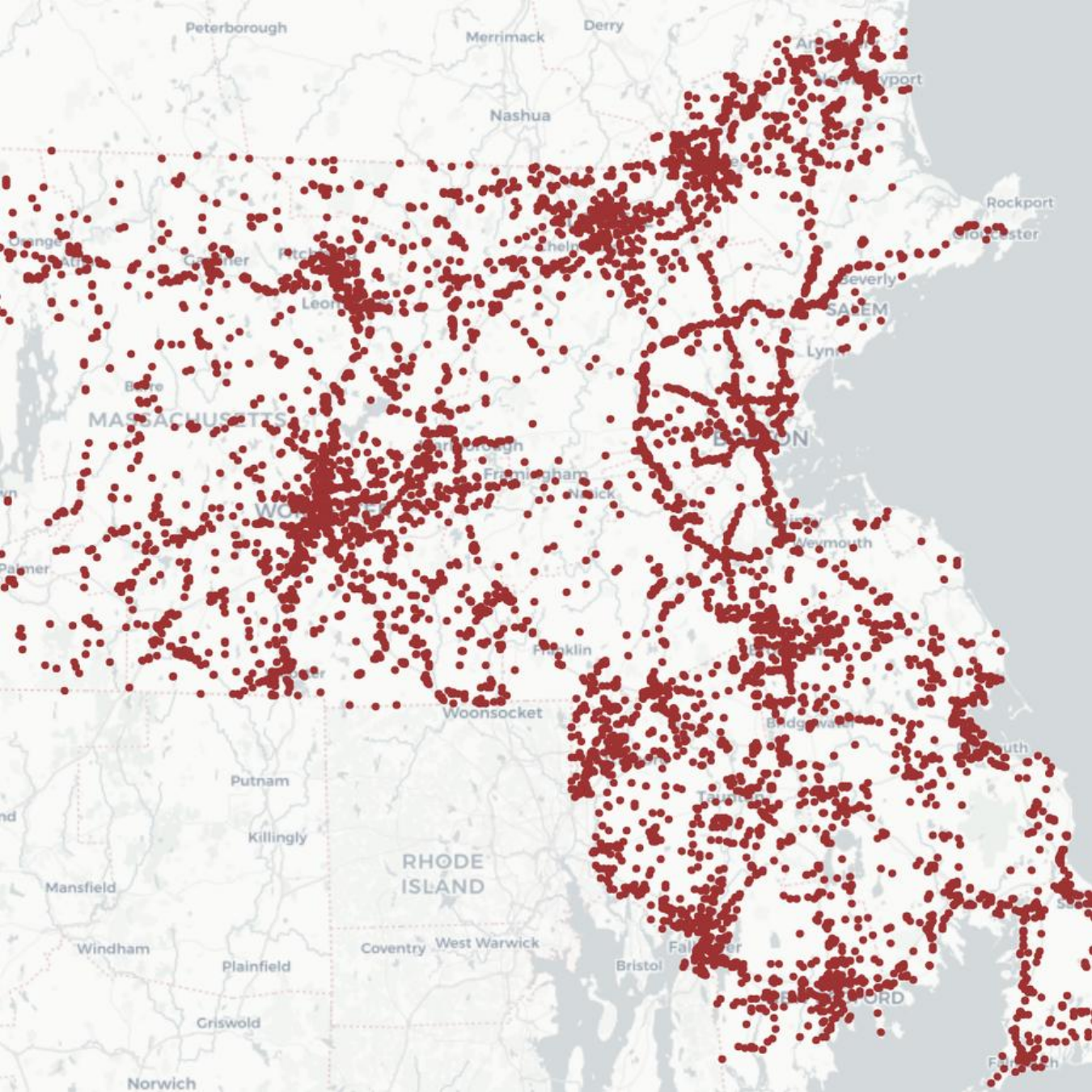}
     \caption{MA, Traffic Volume}
 \end{subfigure}
\end{minipage}\hfill
\begin{minipage}[b]{0.49\textwidth}
  \begin{subfigure}[b]{0.49\textwidth}
     \centering
     \renewcommand\thesubfigure{\alph{subfigure}1}
     \includegraphics[width=\textwidth]{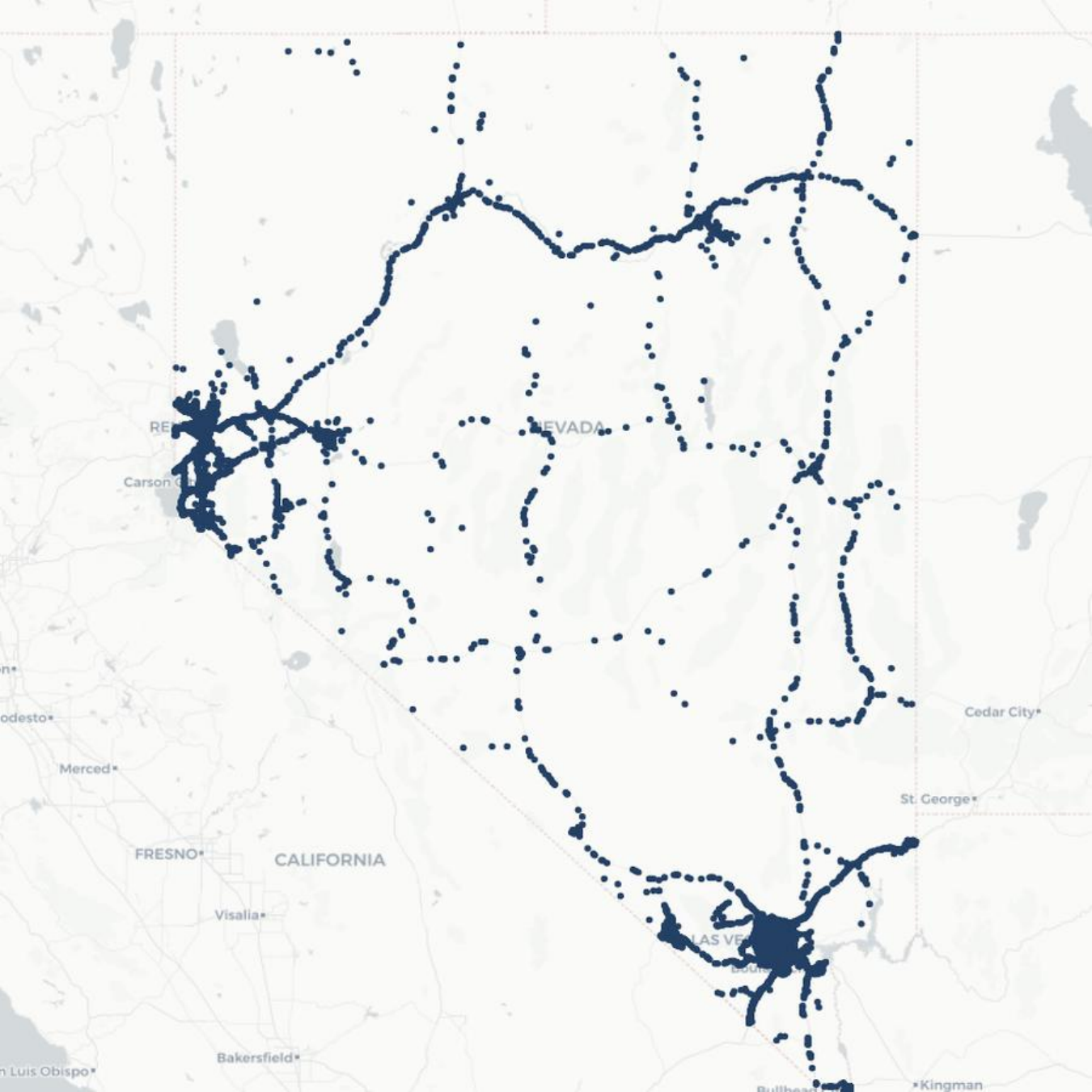}
     \caption{NV, Accidents}
 \end{subfigure}\hspace{0.01in}
 \begin{subfigure}[b]{0.49\textwidth}
     \centering
     \addtocounter{subfigure}{-1}
     \renewcommand\thesubfigure{\alph{subfigure}2}
     \includegraphics[width=\textwidth]{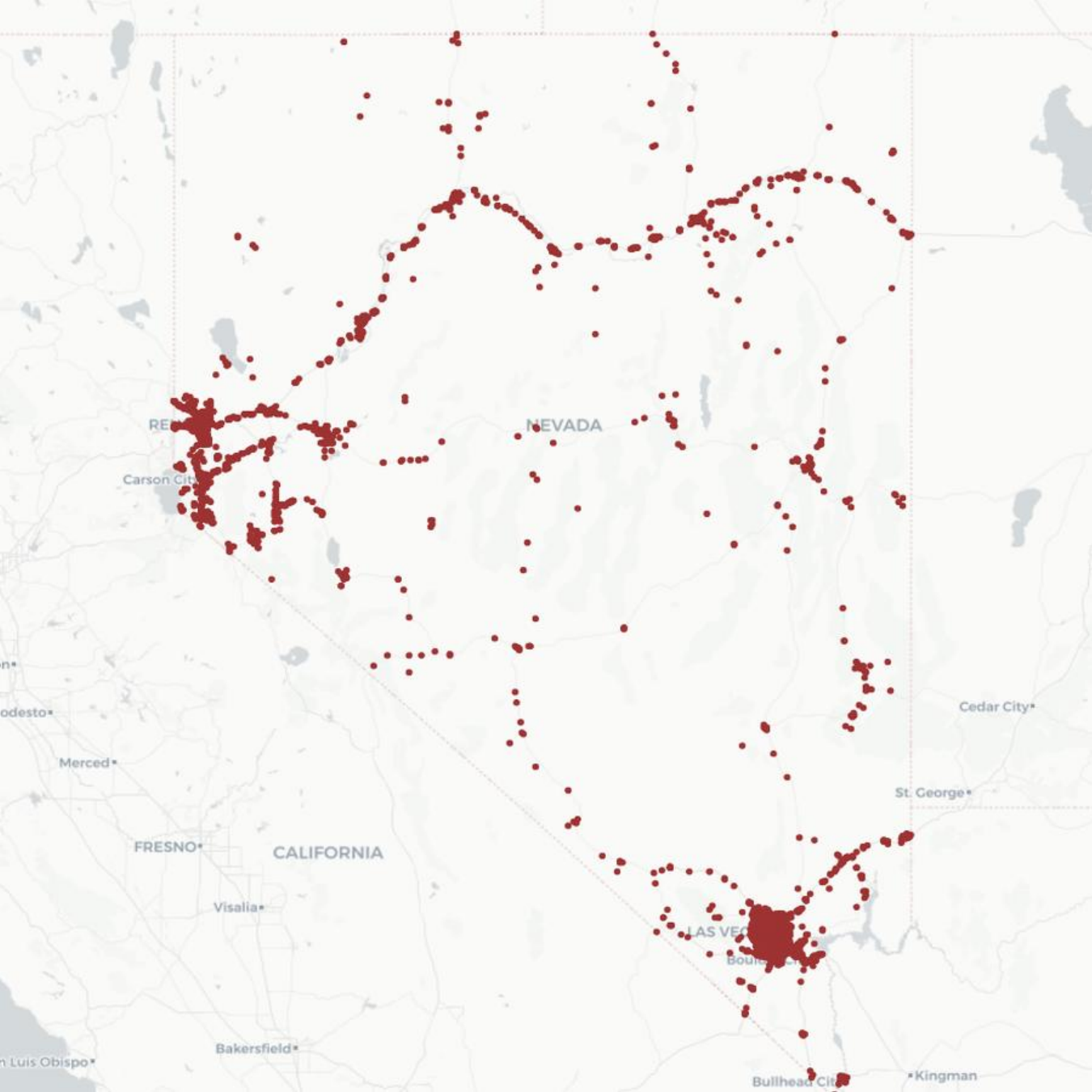}
     \caption{NV, Traffic Volume}
 \end{subfigure}
\end{minipage}
\caption{We note there is a clear association between accident occurrences and traffic volume. Combining accident and annual average daily traffic reports using transfer learning techniques can improve accident prediction by 4.6\%. Further, road network structural features across states are most predictive of accident occurrences. We capture cross-state variability using multitask learning by combining labels of all states. This enables the transfer of information from states with rich data to those with fewer labels. We find that this outperforms learning from individual state data by 4.7\%.}\label{fig_illustration}
\end{figure}

Based on this new dataset, we evaluate existing neural network models in terms of their performance in predicting accident occurrences.
Our major finding is that using road structural features and traffic volume reports, existing graph neural networks such as GraphSAGE \cite{hamilton2017inductive} can predict the accident counts with {\bf 22\%} mean absolute error (relative to actual counts) and whether an accident occurs on the road with over {\bf 87\%} AUROC, averaged over eight states.
We achieve this result by developing multitask and transfer learning techniques on top of the graph neural network, inspired by recent developments in this space \cite{li2023identification,li2023b,ju2023generalization}.
Interestingly, we notice strong cross-sectional trends regarding graph structural patterns across states, as illustrated in Figure \ref{fig_illustration}.
As a remark, we clarify that the results should be interpreted as showing that simple graph neural networks achieve comparable performance to (if not outperforming) their more sophisticated counterparts.
Part of the reason is that our dataset is sparsely labeled, with a labeling rate of less than 0.3\% for all states (cf. Table \ref{tab_accident_binary_classification}), thus making it challenging to fit complex models that have several times more parameters.
One limitation of our analysis is that the road networks, which we extract from OpenStreetMap \cite{haklay2008openstreetmap,boeing2017osmnx}, make simplifying assumptions without considering traffic flows in sequences of connected road segments with typical routes and multi-lanes. We believe our methodology would extend to this case, for example, by combining our accident labels with satellite image-based construction of road networks \cite{he2021inferring}.

To summarize, this paper makes three contributions to traffic accident analysis by learning road network structures.
First, we construct a unified traffic accident dataset from official reports of eight states, totaling nine million records, the largest dataset of this kind to our knowledge.
Second, we find that by using road network structures and traffic flow reports, graph neural networks can accurately predict accident labels, suggesting the validity of our approach.
Third, we discuss the implications of our analysis and develop a package for easily reusing our new datasets and codes.
Our datasets and experiment codes can be found at \url{{https://github.com/VirtuosoResearch/ML4RoadSafety}}.

\section{Methodology}

This section presents our approach to collecting and modeling accident data.
First, we introduce the problem setup.
Second, we describe the collection of a new dataset.
Lastly, we present our modeling approach based on graph neural networks, multitask learning, and transfer learning.

\subsection{Problem setup}\label{sec_setup}

We study the problem of predicting accidents on a road network, viewed as a directed graph $G = (V, E)$, where $V$ denotes a set of road intersections and $E$ denotes a set of roads that connect one intersection to another intersection.
Each node $v \in V$ has a list of node features, including the number of incoming and outgoing edges, its betweenness centrality, and weather information of the corresponding district, among other attributes. 
An edge $e \in E$ has features such as the road length, residential or highway category, and other relevant attributes. 

Each accident is associated with an edge of the road network where the accident happened, as well as the timestamp during which the accident happened. 
Given accident records for a specific time period (e.g., a month), we consider the prediction of accident records for subsequent time periods (e.g., months). %
We measure the result as a regression task by predicting the number of accidents per edge and as a classification task by predicting if one (or more) accidents will occur on a road.

\subsection{Dataset construction}

\textbf{Accident data.} Carrying out machine learning for accident analysis requires collecting historical accident information and predictive features.
There are several widely used traffic network datasets, such as METR-LA and PEMS-Bay \cite{jagadish2014big,li2017diffusion}.
Yet, these curated datasets do not contain vehicle crash information.
There are also online data sources that provide available accident records for the US \cite{kaggle_dataset}.
We note that the dataset is collected from streaming APIs that only provide accident information for certain times of the day (e.g., during rush hours) \cite{moosavi2019accident}.
Further, there is some discussion that the data involves reporting errors in the start and end time \cite{kaggle_dataset}.
Thus, to ensure the validity of our analysis, we start by collecting data from the official reports of the Department of Transportation and note that several states publish detailed information online, including the latitude and longitude of an occurrence.
However, extracting this information is nontrivial: Different states provide the data in different formats and interfaces (some in PDF files).
It is not obvious how to combine all these records in a unified format.
Thus, our first task is to construct a unified dataset with these records.

To this end, we collected over 9 million records spanning eight states in the US.
We provide the basic statistics of this dataset in Table \ref{tab_data}.
We report our statistics at the state level, including the reported number of accidents per million vehicle miles traveled per year, as well as the number of monthly accidents per state. 
However, it is possible to perform this analysis at the county level: Some counties, such as New York City and Los Angeles, report traffic accident information within the county (see links in Table \ref{table_data_source}, Appendix \ref{sec_data_construction}).
For other details of the collection process, see Appendix \ref{sec_data_construction}.

\textbf{Annual average daily traffic (AADT) reports.} Besides, we collect traffic flow records, which are related to crash frequency.
The AADT measures the average number of vehicles traveling on a road per day.
This feature has been known to affect prediction in classical works that apply regression analysis on crash data \cite{persaud1992accident,fridstrom1995measuring,hadi1995estimating,caliendo2007crash}.
However, collecting this data is nontrivial and has not been done in prior works.
We collect official reports of \emph{road-level} AADT from the Department of Transportation, which publishes this information under the name of each street. We map the street names to the edges by extracting a coordinate (using Google Map API) and then aligning it to our graph.

\textbf{Road features.}
For each state, we generate its road network based on OpenStreetMap \cite{haklay2008openstreetmap,boeing2017osmnx}.
Note that this protocol has been widely used in prior traffic forecasting studies (e.g., \citet{li2018multi,geng2019spatiotemporal,he2020roadtagger}).
In Table \ref{tab_data}, we can see that these road networks are very sparse. %
In addition, we have also collected many other features to help with prediction.
For each road, we collect static features including its road category and length information.
There are 24 categories in total, such as one-way, highway, and residential roads (See Appendix \ref{sec_data_construction} for the comprehensive list).
We encode the physical length in meters as a real-valued feature.
For each intersection, we also compute its in-degree, out-degree, and betweenness centrality (which measures the ratio of shortest paths between all node pairs that contain a node), apart from its latitude and longitude.
Next, we collect temporal features of historical (daily) weather information for each node. %
These include the maximum, minimum, and average temperature, the total precipitation of rainfall and snowfall, the average wind speed, and the sea level air pressure.
We align each node with the nearest meteorological station. 

\begin{table}[t!]
\caption{Below are the statistics of collected traffic accidents and features in eight states. We calculate the crash rate as the number of accidents per vehicle mileage traveled (VMT) per year using reported mileage values from the corresponding state.
We use $d_{avg}$ and $d_{max}$ to denote the average and maximum degree of a graph, respectively.
The volume percentage refers to the fraction of roads for which traffic volume reports are available.}\label{tab_data}
\centering
\resizebox{0.8\columnwidth}{!}{
\begin{tabular}{lcccccccc}
\toprule
& Start & End & Crash Rate & \# Nodes & \# Roads & \# Accidents per Month \\ %
\midrule
Delaware & 2009 & 2022 & 3.27 & 49,023  & 116,196 & 2,763 \\ %
Iowa & 2013 & 2022 & 4.92 & 253,623 & 707,072 & 4,451 \\ %
Illinois & 2012 & 2021 & 36.7 & 627,661 & 1,647,614 & 24,839 \\ %
Massachusetts & 2002 & 2023 & 24.48 & 285,942 & 706,402 & 17,723 \\ %
Maryland & 2015 & 2022 & 11.44 & 250,565 & 580,526 & 9,149 \\ %
Minnesota & 2015 & 2023 & 5.39 & 383,086 & 979,259 & 5,711 \\ %
Montana & 2016 & 2020 & 1.69 & 145,525 & 351,516 & 1,665 \\ %
Nevada & 2016 & 2020 & 5.42 & 121,392 & 292,674 & 3,955 \\ %
\bottomrule
\end{tabular}}
\resizebox{0.8\columnwidth}{!}{
\begin{tabular}{lcccccccccc}
\toprule
 & $d_{avg}$  & $d_{max}$ & Centrality ($\times 10^{-3}$) & Avg Length (m) & Volume (\%) \\ %
\midrule
Delaware & 2.4  &  6 & 5.7   & 213 & 3.14 \\ %
Iowa & 2.8 & 7 & 1.4   & 532 & - \\ %
Illinois & 2.6 & 8 & 0.8 & 307 & - \\ %
Massachusetts & 2.5 & 8 & 0.9  & 188 & 1.34 \\ %
Maryland & 2.3 & 8 & 1.0  & 211 & 1.76 \\ %
Minnesota & 2.5 & 8 & 0.6  & 474 & - \\ %
Montana & 2.4 & 7 & 0.02   & 859 & - \\ %
Nevada & 2.4 & 6  & 0.4   & 280 & 1.38 \\ %
\bottomrule
\end{tabular}}
\end{table}

\subsection{Traffic accident prediction}

\textbf{Graph neural networks (GNN).}
The basic unit for our predictive analysis is graph neural networks. Given a graph $G = (V, E)$, along with node-level and edge-level features, a graph neural network applies a neighborhood aggregation mechanism through the edges $E$.
Let $l$ be the number of layers. A GNN recursively computes the representations of a node for $l$ layers by aggregating the representations from its neighbors.
Let $x_i^{(k)}$ denote the node feature at node $i$ in layer $k$ and $v_{i,j}$ denote the features of edge $(i, j)$. 
The $k$-th layer of a GNN aggregates node $i$'s neighborhood embeddings to output:
{\begin{align}
    x_i^{(k+1)} = \phi \left( x_i^{(k)}, h\Big(\Big\{ \psi\big(x_i^{(k)}, x_j^{(k)}, v_{i,j}\big): j\in N(i) \Big\}\Big) \right), \text{ for any } i \in V,
\end{align}}%
where $h$ denotes an aggregation function (e.g., element-wise sum, mean, and max) and $N(i)$ is a set of nodes adjacent to $i$, $\phi$ and $\psi$ denote neural networks with a nonlinear map such as ReLU.

\textbf{Cross-state analysis.} Next, we develop multitask learning (MTL) models to capture cross-state variability.
We note that some states, such as Massachusetts, have way more accident records available than other states, such as Montana.
Thus, combining states with more labels can help predict trends for states with fewer labels.
To ensure that this pooling strategy works, we also require some structural similarities in the feature representations of different states \cite{wu2020understanding}.
 
\begin{figure}[t!]
     \centering
     \begin{subfigure}[b]{0.24\textwidth}
         \centering
         \includegraphics[width=\textwidth]{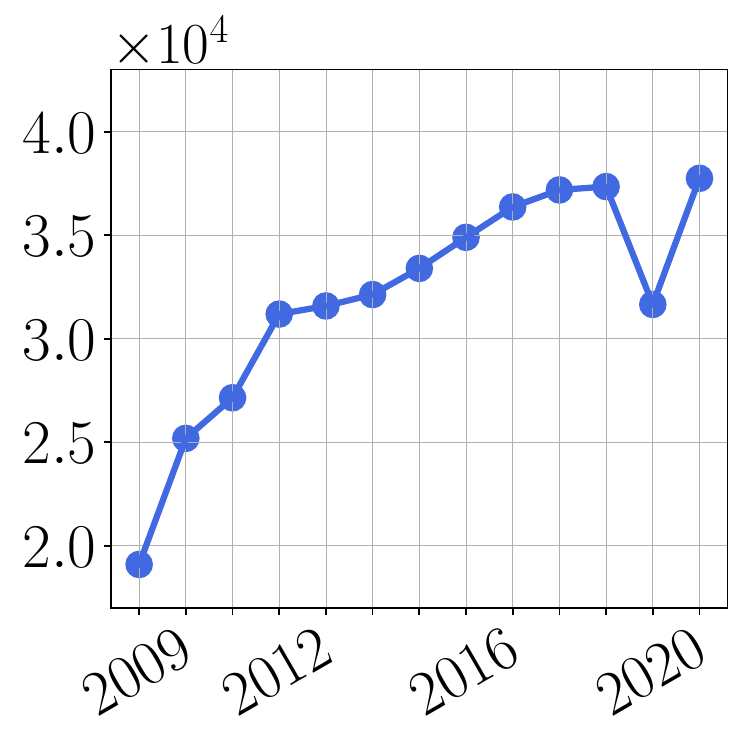}
         \caption{Delaware}\label{fig_delaw}
     \end{subfigure}
     \begin{subfigure}[b]{0.24\textwidth}
         \centering
         \includegraphics[width=\textwidth]{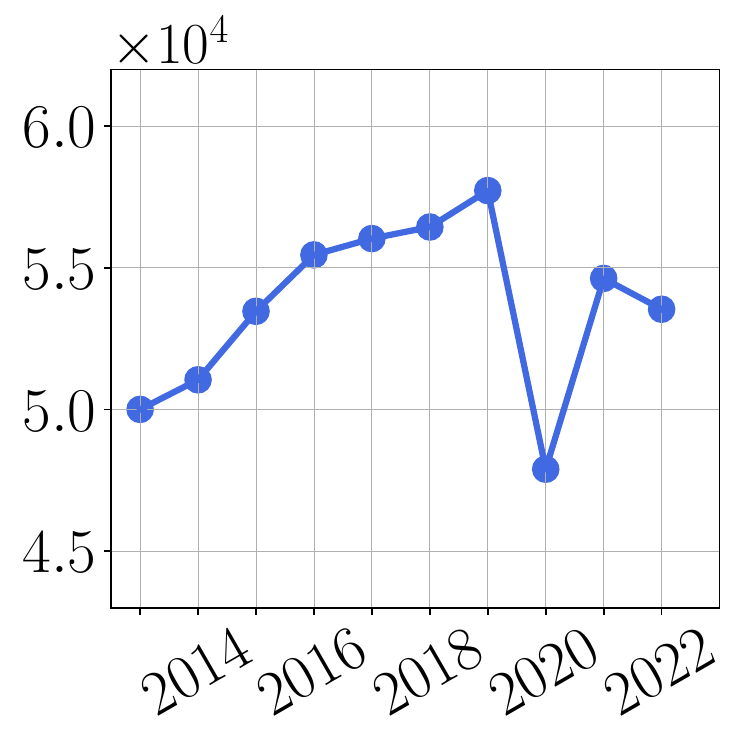}
         \caption{Iowa}
     \end{subfigure}
     \begin{subfigure}[b]{0.24\textwidth}
         \centering
         \includegraphics[width=\textwidth]{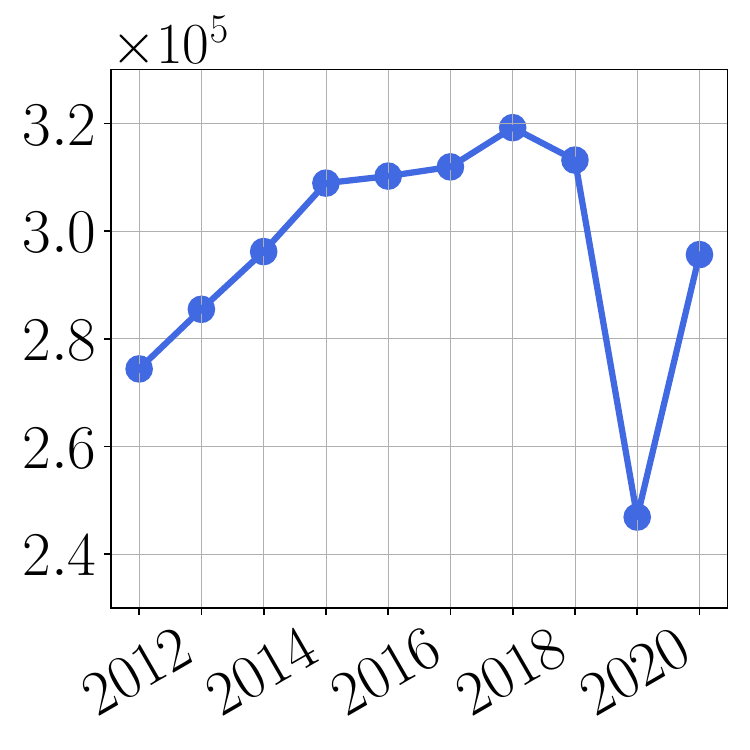}
         \caption{Illinois}
     \end{subfigure}\hfill
     \begin{subfigure}[b]{0.24\textwidth}
         \centering
         \includegraphics[width=\textwidth]{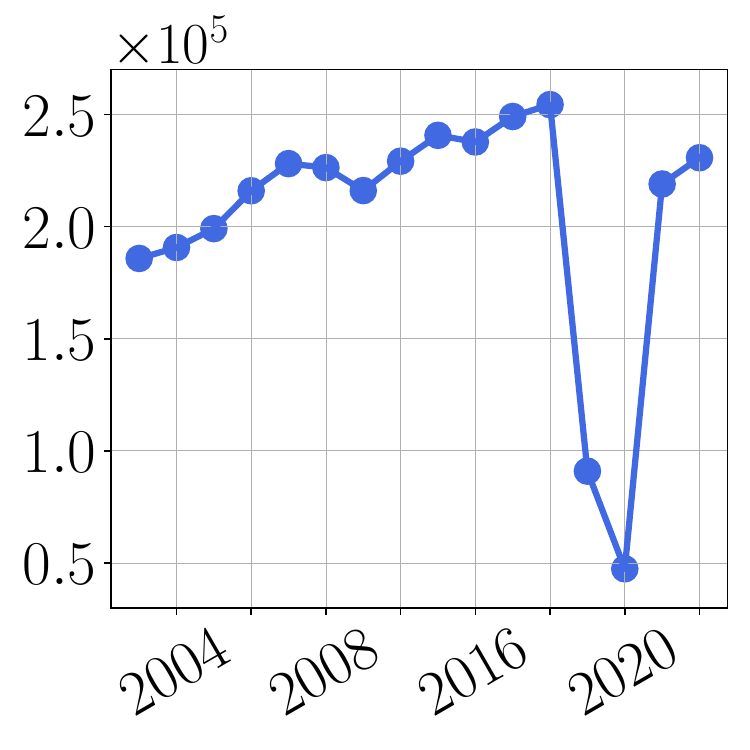}
         \caption{Massachusetts}\label{fig_mass}
     \end{subfigure}
     \centering
    \renewcommand\thesubfigure{\alph{subfigure}1}
     \begin{subfigure}[b]{0.24\textwidth}
         \centering
         \includegraphics[width=0.85\textwidth]{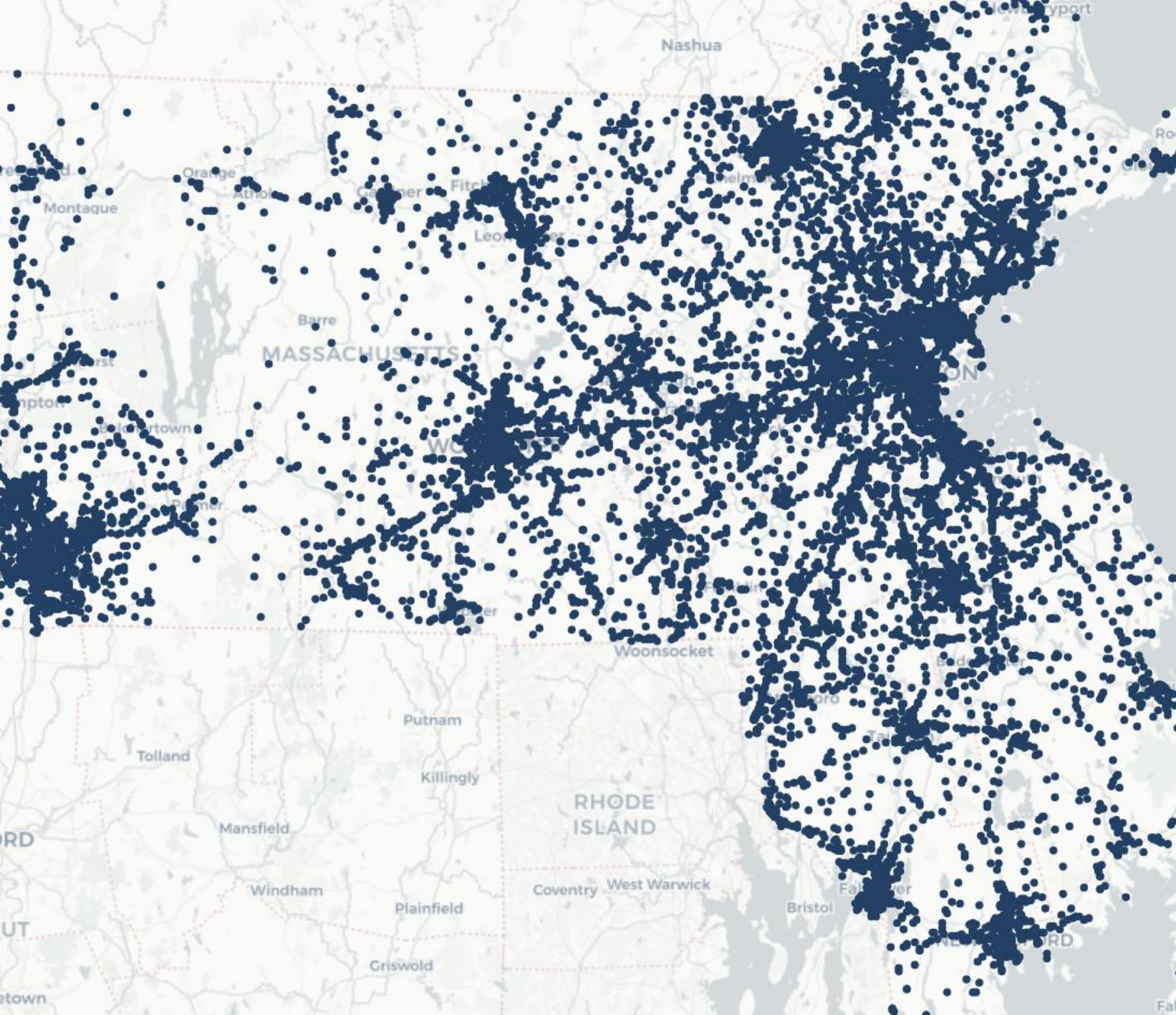}
        \caption{MA, Winter}\label{fig_MA_w}
     \end{subfigure}\hfill
     \begin{subfigure}[b]{0.24\textwidth}
         \centering
        \addtocounter{subfigure}{-1}
        \renewcommand\thesubfigure{\alph{subfigure}2}
         \includegraphics[width=0.85\textwidth]{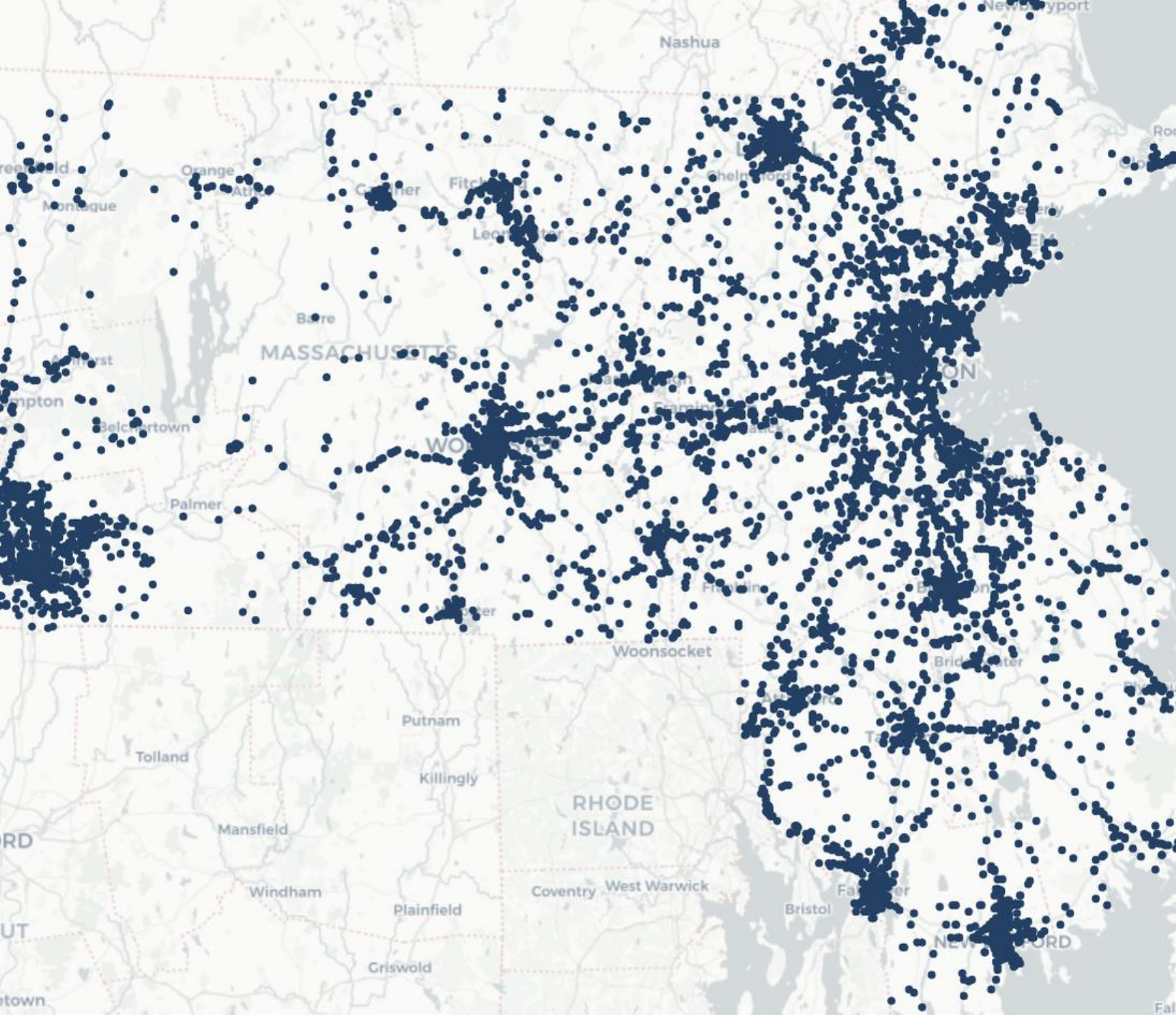}
         \caption{MA, Spring}
     \end{subfigure}\hfill
     \begin{subfigure}[b]{0.24\textwidth}
         \centering
         \renewcommand\thesubfigure{\alph{subfigure}1}
         \includegraphics[width=0.85\textwidth]{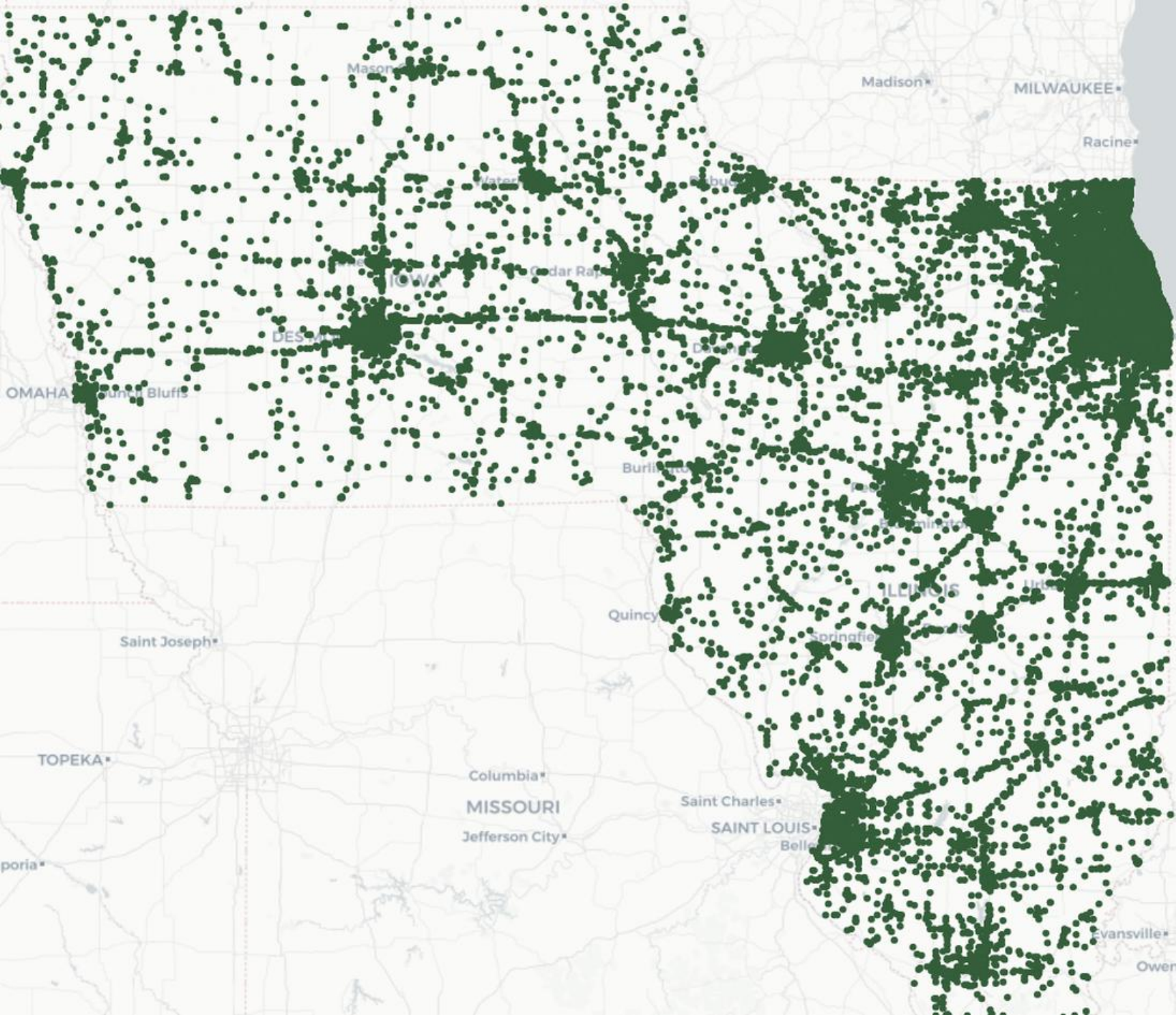}
         \caption{IA \& IL, Winter}
     \end{subfigure}\hfill
     \begin{subfigure}[b]{0.24\textwidth}
         \centering
        \addtocounter{subfigure}{-1}
        \renewcommand\thesubfigure{\alph{subfigure}2}
         \includegraphics[width=0.85\textwidth]{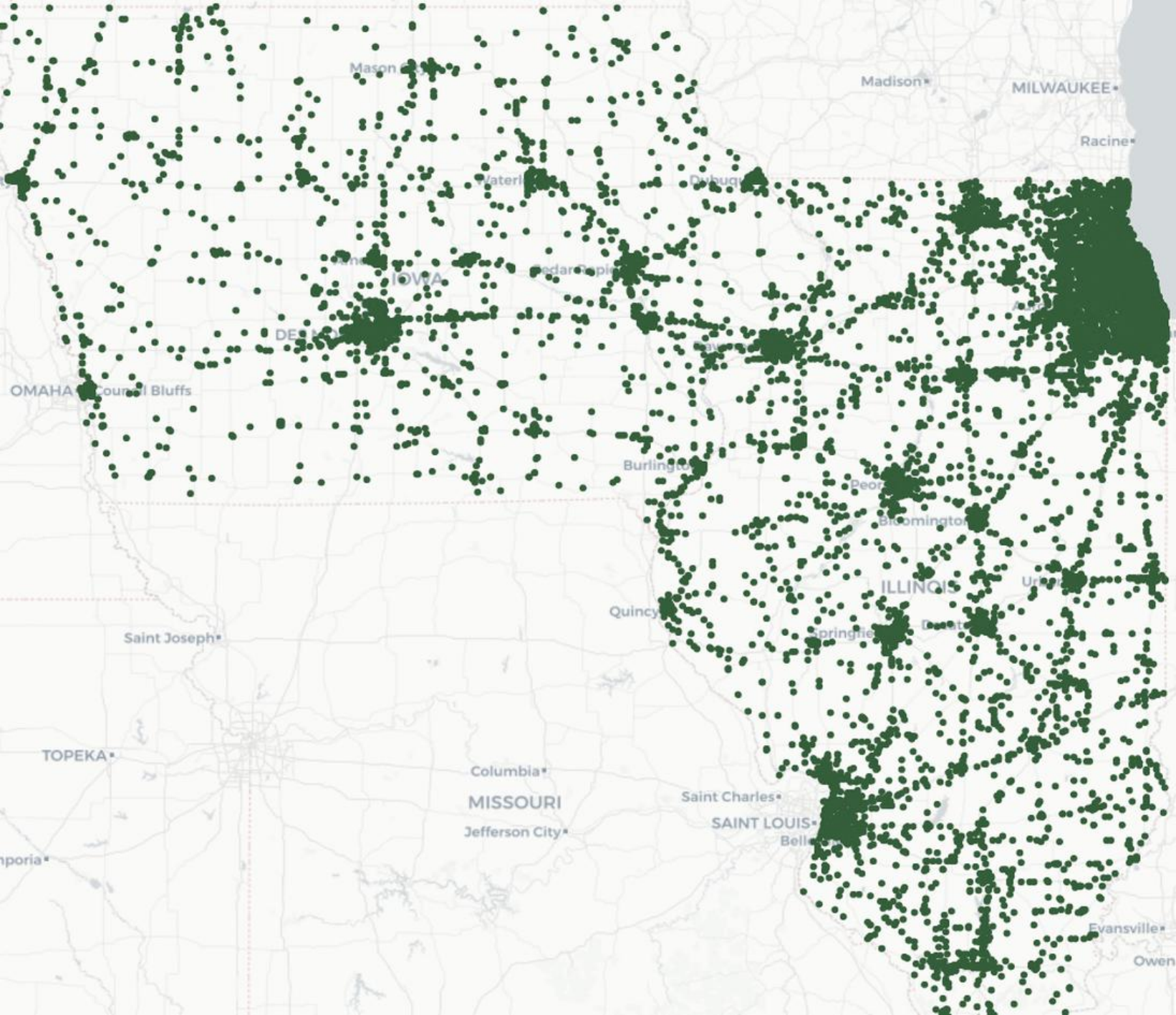}
         \caption{IA \& IL, Spring}\label{fig_ia_sp}
     \end{subfigure}
    \caption{\ref{fig_delaw}-\ref{fig_mass}: We show the evolution of annual accident counts across states. There is a sharp drop in 2020 due to the pandemic.
        \ref{fig_MA_w}-\ref{fig_ia_sp}: We illustrate the seasonal pattern of accidents, where more accidents occur during winter compared to spring.}
        \label{fig_yearly_monthly_accident_trend}%
\end{figure}

Interestingly, we observe cross-sectional trends shared across states.
In Figure \ref{fig_yearly_monthly_accident_trend}, we visualize the number of accidents across four states, including Delaware, Iowa, Illinois, and Massachusetts.
In the top panel, we observe a gradual increase in the number of accidents over the years until 2019, followed by a dip in 2020 due to pandemic-related lockdowns.
Nevertheless, after 2020, the accident count increased again.

Multitask learning encodes such an inductive bias by using a shared encoder for all tasks. %
Let $f(\cdot)$ denote the encoder model.
For each state, we use a separate prediction layer to map the feature vector to an output. Denote these as $h_1(\cdot), h_2(\cdot), \ldots, h_k(\cdot)$.
To train these layers, we combine data from all states and train a multitask model that yields predictions simultaneously for all states.
In particular, we minimize the average loss over the combined dataset of the accident labels of all states.

\textbf{Combining road network geometry and traffic flow.}
Lastly, we develop a transfer learning (TL) technique to combine traffic flow information, which has been shown to relate to the crash frequency in regression models \cite{persaud1992accident}, with network features.
We utilize annual traffic volume reports to support accident prediction by incorporating an additional labeling task into our model.
In this task, we utilize traffic volume information as edge labels alongside the traffic records.
Given traffic volume up to time $t$, we aim to predict traffic volume from $t + 1$ onwards.
This is a regression task at an annual level: Given an edge and a year, the task is to predict the average traffic volume on the road.
Importantly, we exclude the traffic volume feature for this prediction.
Then, we combine this task with accident prediction in a multitask learning model.
We can view accident prediction as the primary task $T_1$ and volume prediction as an auxiliary task $T_2$.
We jointly train a shared encoder $f(\cdot)$ and two separate prediction layers $h_1(\cdot)$ and $h_2(\cdot)$ on the averaged loss of both tasks. 

\begin{wrapfigure}[9]{r}{0.325\textwidth}
     \centering
     \vspace{-0.195in}
     \includegraphics[width=0.30\textwidth]{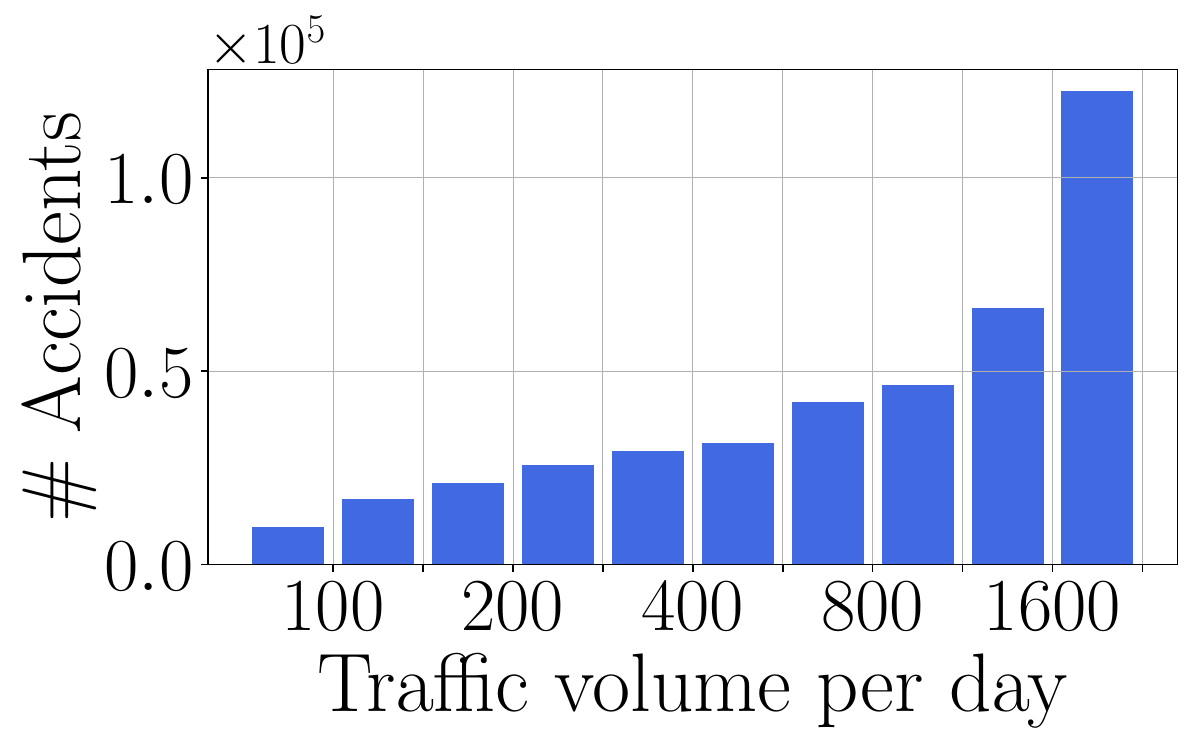}
     \vspace{-0.105in}
     \caption{Distribution of accidents by daily traffic volume.}\label{fig_acc_vs_traffic}
\end{wrapfigure}
We plot the distribution of accidents relative to the corresponding daily traffic volume in Figure \ref{fig_acc_vs_traffic}.
For volume, we gather the average daily traffic records for each road and aggregate the number of accidents at various intervals.
The result indicates a positive correlation between higher traffic volume and increased accidents. 
Consequently, one might expect a positive transfer from the volume prediction task to the accident prediction task when they are trained jointly in a shared model \cite{li2023b}.

\section{Experiments}\label{sec_exp}

This section evaluates various graph learning methods to predict accident labels based on our new dataset.
We focus on three questions.
How effective are existing graph learning methods in capturing the patterns of accidents? 
Does multitask learning capture cross-sectional trends to improve the prediction?
Will traffic volume information help with accident prediction?
We provide positive results for the three questions.
Additionally, we also report ablations highlighting the importance of graph structure and macro-level positive correlations captured by multitask and transfer learning techniques.
Lastly, we discuss the implications of our experimental findings.

\subsection{Experimental setup}

\textbf{Baselines.}
Recall that we consider an edge-level link prediction task.
We use both embedding methods and graph neural networks as baselines.
First, we test multilayer perceptrons (MLP) using the node features, including node degrees, betweenness centrality, and weather. This tests the performance of using node features without network structures.
Second, we test embedding methods, including Node2Vec \cite{grover2016node2vec} and DeepWalk \cite{perozzi2014deepwalk}, and add a layer to concatenate the node embeddings and features.
Third, we evaluate various GNN architectures, including GCN \cite{kipf2016semi}, GraphSAGE \cite{hamilton2017inductive}, and GIN. %
Lastly, we also evaluate spatiotemporal GNNs that model the temporal dependency of the time series features. 
These include DCRNN \cite{li2017diffusion}, STGCN \cite{yu2017spatio}, AGCRN \cite{bai2020adaptive}, and Graph Wavenet \cite{wu2019graph}.
To enhance the prediction of GNN, we add Node2Vec embeddings as node features.

We consider three methods to conduct multitask and transfer learning: 
First, we fit multitask GNNs by combining the data of all eight states. We denote this model as MTL.
Second, after fitting an MTL model, we fine-tune the feature encoder on an individual state's data (or MTL-FT in short).
Third, we train a model on accident and volume prediction jointly, denoted as TL.
For all the methods above, we use GraphSAGE as the feature encoder. 

\textbf{Implementations.} 
For all baselines, we construct the feature representation for each edge by concatenating the encoded representations of two adjacent nodes and the edge-level features. 
In our implementation, we fix the dimension of node embeddings to 128.
We use a two-layer MLP and GNN with a hidden dimensionality of 256. 
We train our models using Adam as the optimizer.
We use a learning rate of 0.001 for 100 epochs on all models.
These hyperparameters are tuned with grid search on the validation set for all the models. 
The number of layers is tuned in a range of $\set{2, 3, 4}$.
The hidden dimensionality is tuned in a range of $\set{128, 256, 512}$. 
The learning rate is tuned in a range of $\set{0.01, 0.001, 0.0001}$. 
The number of epochs is tuned in a range of $\set{50, 100, 200}$. 
See Appendix \ref{sec_detailed_exp} for more implementation details. 

For each state, we evenly split the available period of accidents into training, validation, and test sets.
We split the accidents according to time. We use past accident records until a specific year to train the models and evaluate the model's performance on future accidents occurring after that year. %
We focus our prediction monthly by summing up the daily accident occurrences into a monthly count, but note that one can use our datasets to conduct the analysis at the daily or the annual level too.
We use both regression and classification metrics to evaluate the results.
For regression tasks, we measure the mean absolute error (MAE) between the predicted number of accidents and the actual number of occurrences on a particular road.
For classification tasks, we measure the AUROC scores. %

\subsection{Experimental results}\label{sec_exp_results}

We summarize our experimental results in Table \ref{tab_accident_binary_classification}.
We highlight three conceptual takeaways below, which we believe will also apply more broadly beyond the specific setting of our experiment.

\textbf{(1) Graph neural networks can accurately predict accident labels.}
We find that using graph neural networks can predict accident counts with \textbf{0.3} mean absolute error, which is \textbf{22\%} relative to the absolute accident count on average over eight states.
For classifying whether an accident occurs or not, GNNs can achieve \textbf{87\%} AUROC score on average.

Among the GNNs, we observe that the comparisons between GraphSAGE and other spatiotemporal GNNs are generally mixed, with no single architecture dominating the others. While GraphSAGE has fewer parameters, its performance is still comparable with (if not outperforming) alternative models with two times or more parameters. One explanation is that spatiotemporal GNNs have more trainable parameters than GraphSAGE. Therefore, they need more training labels to fit the model. On the other hand, the labeling rate of our dataset, i.e., the percentage of edges with a positive label or accident occurrence, is around or below 0.2\%, as shown in Table \ref{tab_accident_binary_classification} (under the row of ``Positive Rate'').

\textbf{(2) Multitask learning captures macro-level trends across states.} We also find that multitask learning outperforms single-task learning (STL) by relatively \textbf{8.4\%} in terms of MAE and \textbf{0.9\%} in terms of AUROC averaged over eight states.
This is achieved by first training a model on the combined data of all states and then fine-tuning the MTL model on each state's data.

\textbf{(3) Transfer learning from traffic volume prediction improves test performance.}
Lastly, we find that combining traffic volume and accident prediction yields a relative improvement of \textbf{7.9\%} in MAE and \textbf{1.1\%} in AUROC over STL, averaged over the four states with traffic volume records.

\begin{table}[t!]
\centering
\caption{We compare the experimental results across eight states using node embedding methods and graph neural networks. We also include multitask and transfer learning results for each state. We report the mean absolute errors (MAE) in predicting the accident counts on the test split. We also report the AUROC score in predicting the accident occurrences on the test split. To measure standard deviations, we run the same experiment over three random seeds and report the average result.} \label{tab_accident_binary_classification}
\resizebox{\columnwidth}{!}{ %
\begin{tabular}{@{}lcccccccc@{}}
\toprule
MAE ($\downarrow$) & DE & IA & IL & MA & MD & MN & MT & NV \\
Avg Count & 1.23 & 1.14 & 1.33 & 2.27 & 1.22 & 1.18 & 1.16 & 1.38 \\
\midrule
MLP & 1.4$\pm$0.07 & 0.3$\pm$0.02 & 1.4$\pm$0.17 & 1.0$\pm$0.11 & 0.4$\pm$0.02 & 0.3$\pm$0.02 & 0.4$\pm$0.01 & 0.5$\pm$0.01 \\
Node2Vec & 1.1$\pm$0.18 & 0.3$\pm$0.01 & 0.7$\pm$0.05 & 1.3$\pm$0.51 & 0.4$\pm$0.03 & 0.4$\pm$0.02 & 0.2$\pm$0.03 & 0.3$\pm$0.01  \\
DeepWalk & 0.8$\pm$0.05 & 0.3$\pm$0.01 & 0.6$\pm$0.03 & 1.0$\pm$0.06 & 0.4$\pm$0.01 & 0.4$\pm$0.01 & 0.3$\pm$0.03 & 0.3$\pm$0.02\\
GCN & 0.6$\pm$0.02 & 0.3$\pm$0.02 & 0.5$\pm$0.06  & {0.7$\pm$0.02} & {0.4$\pm$0.04} & 0.3$\pm$0.00 & 0.2$\pm$0.03 & 0.3$\pm$0.02\\ 
GraphSAGE & {0.3$\pm$0.01} & {0.3$\pm$0.01} & 0.4$\pm$0.03  & 0.8$\pm$0.02  & {0.4$\pm$0.01} & 0.3$\pm$0.01 & {0.2$\pm$0.02} & {0.2$\pm$0.01} \\
GIN & 0.8$\pm$0.02 & 0.3$\pm$0.04 & {0.4$\pm$0.04} & 1.1$\pm$0.02 & 0.4$\pm$0.02 &  {0.3$\pm$0.06} &  0.2$\pm$0.08 & 0.2$\pm$0.05 \\
{AGCRN} & 0.3$\pm$0.01 & 0.3$\pm$0.01 & 0.4$\pm$0.01 & 0.7$\pm$0.01 & 0.4$\pm$0.01 & 0.3$\pm$0.01 & 0.2$\pm$0.01 & 0.2$\pm$0.04 \\
{STGCN} & 0.2$\pm$0.02 & 0.3$\pm$0.00 & 0.4$\pm$0.06 & 0.8$\pm$0.06 & 0.5$\pm$0.01 & 0.3$\pm$0.01 & 0.2$\pm$0.01 & 0.2$\pm$0.01 \\
{Graph Wavenet} & 0.3$\pm$0.03 & 0.3$\pm$0.02 & 0.4$\pm$0.00 & 0.7$\pm$0.01 & 0.3$\pm$0.00 & 0.3$\pm$0.01 & 0.3$\pm$0.03 & 0.2$\pm$0.01 \\
DCRNN  & 0.3$\pm$0.01 & 0.3$\pm$0.02 & 0.4$\pm$0.03 & 0.9$\pm$0.06 & 0.3$\pm$0.01 & 0.4$\pm$0.02 & 0.2$\pm$0.02 & 0.2$\pm$0.03\\
\midrule
MTL & 0.2$\pm$0.02 & 0.2$\pm$0.01 & 0.4$\pm$0.02 & 0.7$\pm$0.02 & 0.3$\pm$0.00 & 0.2$\pm$0.01 & 0.1$\pm$0.00 & 0.2$\pm$0.01\\
MTL-FT & \textbf{0.2}$\pm$0.01 & \textbf{0.2}$\pm$0.00 & \textbf{0.2}$\pm$0.00 & 0.7$\pm$0.01 & \textbf{0.3}$\pm$0.00 & \textbf{0.2}$\pm$0.01 & \textbf{0.1}$\pm$0.00 & \textbf{0.2}$\pm$0.00\\
TL & 0.2$\pm$0.01 & - & - & \textbf{0.6}$\pm$0.02 & 0.3$\pm$0.01 & - & - & 0.2$\pm$0.01 \\
\midrule\midrule
AUROC ($\uparrow$) & DE & IA & IL & MA & MD & MN & MT & NV \\
Training Size & 93,184 & 187,046 & 646,739 & 540,682 & 283,226 & 124,435 & 34,475 & 73,164\\
Positive Rate & 0.23 & 0.07 & 0.14 & 0.10 & 0.15 & 0.05 & 0.05 &  0.12 \\
\midrule
MLP & 75.5$\pm$0.6 & 68.7$\pm$0.1 & 71.4$\pm$0.3 & 70.6$\pm$0.2 & 74.1$\pm$0.4 & 76.7$\pm$0.2 & 71.6$\pm$0.1 & 56.2$\pm$0.1\\ %
Node2Vec & 83.5$\pm$0.1 & 83.8$\pm$0.1 & 77.4$\pm$1.5 & 70.7$\pm$0.4 & 83.5$\pm$0.1 & 80.6$\pm$0.1 & 84.9$\pm$0.1 & 91.8$\pm$0.1 \\ %
DeepWalk & 83.4$\pm$0.3 & 81.6$\pm$0.2 & 78.6$\pm$0.2 & 69.5$\pm$0.2 & 83.7$\pm$0.1 & 80.5$\pm$0.2 & 85.0$\pm$0.1 & 91.8$\pm$0.1\\ %
GCN & 83.2$\pm$0.1 & {85.4$\pm$0.1} & 84.7$\pm$1.0 & 70.6$\pm$0.1 & 83.2$\pm$0.1  & 84.3$\pm$2.2 & 87.4$\pm$0.1 & {91.9$\pm$0.6} \\ %
GraphSAGE & {87.6$\pm$0.1} & 84.8$\pm$0.2 & {87.0$\pm$0.2} & {81.8$\pm$0.1} & {87.6$\pm$0.1} & 83.8$\pm$2.8 & {87.5$\pm$0.1} & 91.6$\pm$1.0 \\ %
GIN & 82.6$\pm$0.7 & 83.5$\pm$0.1 & 84.2$\pm$1.4 & 68.9$\pm$0.5 & 82.6$\pm$0.7 & {85.4$\pm$1.4} & 85.4$\pm$0.1 & 91.4$\pm$1.0 \\ %
{AGCRN} & 86.0$\pm$0.2 & 83.9$\pm$0.2 &  {86.3$\pm$0.3} & 82.1$\pm$0.2 & 88.5$\pm$0.1 & 81.8$\pm$0.7 & 84.3$\pm$0.4 & 90.7$\pm$0.2\\ %
{STGCN} & 85.4$\pm$0.1 & 83.5$\pm$0.1 & 85.2$\pm$0.4 & 81.9$\pm$0.3 & 88.7$\pm$0.1 & 81.5$\pm$0.2 & 83.8$\pm$0.3 & 91.5$\pm$0.3\\ %
{Graph Wavenet} & 85.0$\pm$0.2 & 83.9$\pm$0.2 & 85.8$\pm$0.2 & 81.9$\pm$0.5 & 87.9$\pm$0.1 & 80.3$\pm$0.1 & 83.4$\pm$0.2 & 90.6$\pm$0.2 \\ %
DCRNN & 81.2$\pm$1.2 & 81.8$\pm$0.1 & 80.7$\pm$0.0 & 70.5$\pm$0.1 & 84.5$\pm$0.3 & 79.3$\pm$0.4 & 81.9$\pm$0.5 & 90.5$\pm$0.7 \\ %
\midrule
MTL & 87.7$\pm$0.1 & 81.7$\pm$0.2 & 84.4$\pm$0.3 & 79.6$\pm$0.1 & \textbf{88.7}$\pm$0.1 & \textbf{87.9}$\pm$0.0 & 88.4$\pm$0.2 & 90.3$\pm$0.2  \\
MTL-FT & \textbf{87.8}$\pm$0.3 & \textbf{84.9}$\pm$0.2 & \textbf{87.2}$\pm$0.2 & {81.9}$\pm$0.3 & 88.1$\pm$0.1 & 87.6$\pm$0.3 & \textbf{88.5}$\pm$0.3 & {91.8}$\pm$0.2 \\
TL & 87.3$\pm$0.2 & - & - & \textbf{82.6}$\pm$0.2 & 87.9$\pm$0.4 & - & - & \textbf{92.8}$\pm$0.1 \\
\bottomrule
\end{tabular} 
}
\end{table}

\subsection{Ablation studies}

\textbf{Influence of graph-structural features.}
We conduct a leave-one-out analysis to assess the importance of different feature categories for accident prediction.
These include graph-structural features, weather, and traffic volume.
We remove one type of feature at a time and compare the performance after leaving out a particular feature.
Removing graph-structural features reduces the performance by 6.9\%.
On the other hand, removing weather and traffic volume reduces performance by 2.3\% and 1.2\%, respectively.
See Table \ref{tab_ablating_features} for the results, which justify that graph-structural features are the most significant features.

\medskip

\begin{wrapfigure}[11]{r}{0.40\textwidth}
     \centering
     \vspace{-0.2in}
     \includegraphics[width=0.28\textwidth]{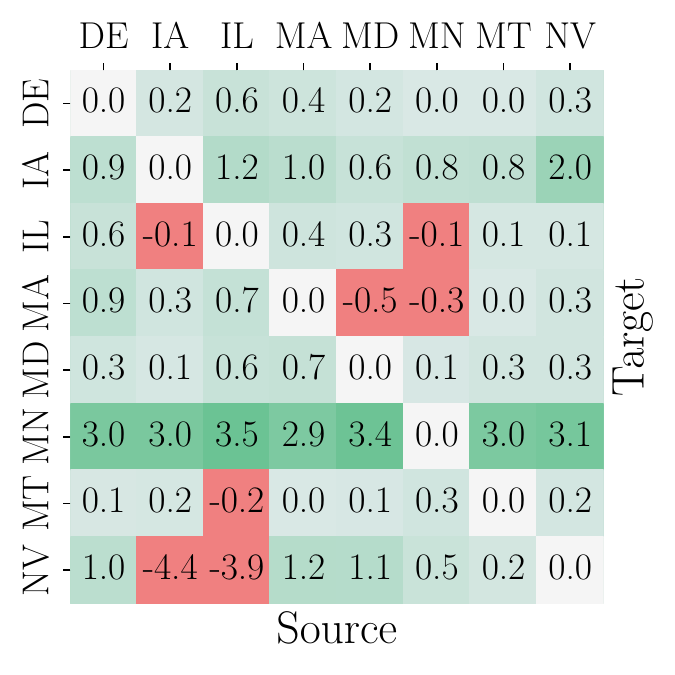}
     \vspace{-0.1in}
     \caption{Pairwise training vs. single-task learning.}\label{fig_pairwise_transfer}
\end{wrapfigure}
\textbf{Positive pairwise transfer across states.}
Next, we measure the transfer effects between every pair of states.
For each state, we view it as a source state and consider how well it would transfer to another target state.
To test this, we conduct MTL by combining each state with another state's data.
This leads to a total of 28 pairwise MTL models.
We show the results in Figure \ref{fig_pairwise_transfer} on the right.
To help read this table, we subtract each MTL model's performance for a target state from the STL performance of that target state.
Thus, a positive value in the table indicates a positive transfer from the source state to the target state.
We find that for most pairs, the effect is positive.

\medskip
\textbf{Transferability from traffic volume to accident prediction.} We also measure the transfer effect from volume prediction to accident prediction.
We first fit a model on the volume prediction task for one state, such as the MA.
Then, we fine-tune this model on the accident prediction task using the accident labels while keeping the same network and other features.
In Table \ref{tab_ablating_features}, this fine-tuned model outperforms training a randomly initialized model to predict the accident labels by 0.6\%, averaging over four states with volume labels.

\begin{table}[t!]
\centering
\caption{We evaluate the influence of graph-structural structure, weather, and traffic volume information. We report the test AUROC of accident classification by removing each feature type in the network. We also include the transfer results of fine-tuning a model trained on traffic volume prediction to accident prediction.} \label{tab_ablating_features}
\begin{small}
\begin{tabular}{@{}lcccccccc@{}}
\toprule
 & DE & MA & MD & NV \\ %
\midrule
Using all features &  87.61$\pm$0.10 & 81.80$\pm$0.12 & 87.51$\pm$0.02 & 91.62$\pm$0.99 \\
w/o graph structural features &  81.25$\pm$0.51 & 79.63$\pm$0.18 & 79.77$\pm$0.94 & 82.56$\pm$0.09 \\ 
w/o weather information & 87.27$\pm$0.32 & 80.71$\pm$0.26 & 80.22$\pm$0.45 & 90.38$\pm$0.88 \\
w/o road information & 82.99$\pm$0.54 & 81.65$\pm$0.77 & 80.63$\pm$0.52 & 84.24$\pm$0.41\\ 
w/o traffic volume information & 87.15$\pm$0.49 & 80.94$\pm$0.31 & 86.17$\pm$1.15 & 91.58$\pm$0.99 \\% 
\midrule
Transfer learning from traffic volume prediction & 87.78$\pm$0.07 & 82.18$\pm$0.14 & 87.77$\pm$0.24 & 92.07$\pm$0.57 \\
\bottomrule 
\end{tabular}
\end{small}
\end{table}

\textbf{Sensitivity analysis.} Lastly, we study the hyperparameters used in our experiments.
In particular, we set the number of layers to $2$, the hidden dimensionality to $256$, the learning rate to $1e^{-3}$, and the number of epochs to $100$.
We then vary one hyper-parameter at a time and keep the others unchanged. 
We notice that using the number of layers as $2$, hidden dimensionality as $256$, and learning rate as $1e^{-3}$ yields the best results for all baselines.
The validation performance stops improving after training up to $100$ epochs. 
Thus, we adopt these settings as the default parameters.

\subsection{Interpretations and implications}

Our findings provide some evidence to show that the road network structure is highly predictive of traffic accident occurrences.
The evaluation of numerous graph learning methods shows that these methods are valid for accident analysis.
Our findings about the road structures go beyond existing regression analysis in the transportation literature (e.g., \cite{persaud1992accident,caliendo2007crash,lin2015novel}) thanks to the collection of a large-scale dataset.
Based on the findings, we discuss several implications for when our datasets might be helpful.

\emph{Studying the effect of policy interventions:}
One way to use our dataset is for policy interventions by examining the accident patterns before and after the implementation of this policy.
Our modeling approach could be useful in policy-making, such as the abolition of mandatory vehicle inspection.
The model predictions can provide counterfactual comparisons to facilitate such discussions.

\emph{Variability at the county level:} With our dataset, it is also possible to study road structures across counties within the same state, which controls for variability with weather conditions. We believe the comparison can also inform network design.

\emph{Better reporting of traffic volume information:}
We believe that better reporting of traffic volume information could be very useful for accident analysis.
Our dataset shows that the percentage of roads where volume reports are available is pretty low (See Table \ref{tab_data}).
For example, we noticed that the accident numbers are high on some roads, but the traffic volume is either low or missing.
Besides, we also note that there is likely under-reporting of accidents (e.g., to reduce insurance costs).
Thus, better traffic flow reports combined with better modeling can help address such questions.

\section{Related Work}

Datasets and benchmarks are instrumental to machine learning research. 
Motivated by developments in machine learning on graphs, several large-scale graph learning benchmarks have been developed \cite{hu2020open,morris2020tudataset,freitas2020large,hu2021ogb}. Examples include a large-scale graph dataset based on Amazon’s product co-purchase network \cite{chiang2019cluster} and a number of graph datasets in the biological domain. MoleculeNet \cite{wu2018moleculenet} introduces a large-scale benchmark for studying molecule graphs and structures. TAPE \cite{rao2019evaluating} describes a set of five biologically relevant semi-supervised learning tasks spread across different domains of protein biology.
There have also been recent developments in three-dimensional representations of molecules and whole-brain vessel graphs based on whole, segmented murine brain images.
TUDataset \cite{morris2020tudataset} consists of over 120 datasets of varying graph sizes from various applications, such as point clouds and social networks.

To facilitate comparison between methods, the open graph benchmark database \cite{hu2020open,hu2021ogb,freitas2020large} provides diverse datasets and unified evaluation protocols across academic collaboration networks, protein structural graphs, and Reddit social networks. 
Besides supervised learning, recent work has developed benchmarks for graph contrastive learning and node outlier detection.
Compared to existing datasets, our work addresses a problem of societal relevance, providing spatiotemporal patterns related to road networks across states.

\begin{table}[t!]
\centering
\caption{Comparison between our dataset vs. several existing accident record datasets.} \label{tab_dataset_comp}
{
\begin{small}
\begin{tabular}{@{}p{1.0 cm} p{3.5cm} p{4.0cm} p{4.0cm} @{}}
\toprule
 & \citet{huang2023tap} & \citet{yuan2018hetero} & Our Dataset \\
\midrule
Data Source & Based on work by Moosavi et al. (2019), sourced from Microsoft Bing Map Traffic & Department of Transportation’s official accident reports & Department of Transportation’s official accident reports\\
\midrule
Coverage & 2.8 million records across 13 states for 2016-2021 & Iowa, for 2006-2013 & 9 million records, across 8 states, for a maximum of 20 years up to 2023\\
\midrule
Prediction  Tasks & Node-level classification, evaluated on AUROC & Spatial grid-level regression, evaluated under MAE & Edge-level regression/classification, evaluated under MAE/AUROC\\
\midrule
Features & Road network features & Averaged grid-level features, traffic volume, rainfall features & Traffic volume, weather features, road network features\\
\bottomrule
\end{tabular}
\end{small}
}
\end{table}

\medskip
\textbf{Spatiotemporal graph mining.}
The importance of spatial and temporal features for time series prediction is well recognized \cite{li2017diffusion,li2018multi}.
Convolutional networks are often used to process time series graphs \cite{yu2017spatio}.
More recently, graph neural networks have been designed for spatiotemporal prediction with uncertainty quantification \cite{zhuang2022uncertainty}. 
Besides model architecture design, multitask and meta-learning techniques are used to tackle spatiotemporal heterogeneity \cite{yao2019learning,li2018multi}.
The transferability of graph representations has been studied for airport networks and gene interaction graphs \cite{zhu2021transfer}.
Our results suggest that road network structures are highly transferable in predicting traffic accidents.
Furthermore, contrastive learning has been extensively explored on graph-structured data \cite{you2020graph,you2021graph}.
Recent studies \cite{qu2022forecasting,zhang2023automated} have formulated contrastive learning methods on spatiotemporal graph data, such as predicting urban flow, crime, and house prices.
One interesting question is to revisit these techniques within the context of traffic accident prediction for road safety.

\medskip
\textbf{Data-driven traffic forecasting.}
There is an extensive literature on traffic network analysis and traffic forecasting \cite{jiang2022graph}.
Existing datasets such as METR-LA and PEMS-BAY are widely used in training machine learning models for traffic forecasting.
These datasets have been collected from loop detectors by California's Transportation Agencies.
Many research studies use taxi data,  such as the NYC open dataset and Didi Chuxing traffic information. One related task is to predict the ride-hailing demand in online ride-hailing platforms \cite{geng2019spatiotemporal}.
There are also studies on predicting the estimated time of arrival \cite{hong2020heteta}, traffic time and distance for a taxi trip \cite{li2018multi}, and traffic speed.
Besides graph structural features, recent work has studied constructing road network features from satellite images \cite{he2020roadtagger}.
Further studying the use of satellite images for large-scale traffic accident prediction is a promising direction for future work.

\medskip
\textbf{Comparison with existing traffic accident prediction datasets.} Several prior works \cite{huang2023tap,yuan2018hetero} have examined large-scale traffic accident prediction by using OpenStreetMaps to construct road network features, including road network features such as the length of a road and the type of road (e.g., highway or residential, one-way).
The difference lies in how we construct the accident records. First, TAP \cite{huang2023tap} used the accident information collected from another work by \citet{moosavi2019accident}, sourced from Microsoft Bing Map Traffic. Their dataset includes a total of 2.8 million accident records, which are for the five years between 2016 and 2021. In contrast, our datasets are collected from official accident reports of the Department of Transportation. Our dataset includes 9 million records for a maximum of 20 years (e.g., Massachusetts, from 2002 to 2023, which also contains more recent data).
Second, each accident record contains the latitude and longitude of the incident. We map each coordinate to the nearest edge, leading to an edge-level classification/regression problem, whereas TAP maps each record to the nearest node, resulting in a node-level classification setting. Thus, our work can be evaluated under both MAE/AUROC metrics.
Another major difference is that we have collected traffic volume and weather features alongside the road network features.
In summary, we give a detailed comparison between our dataset and two existing datasets in Table \ref{tab_dataset_comp}.

\medskip
\textbf{Comparison with existing spatiotemporal datasets.}
Next, we compare several spatiotemporal datasets with ours.
We note that existing datasets focus on rather different tasks, such as predicting traffic speed and volume collected from highway loop detections \cite{li2017diffusion}. Some other works study predicting traffic information from ride-hailing platforms, such as the ride number in a certain region \cite{sun2020predicting} and the estimated trip time \cite{li2018multi}. In contrast, our work aims to predict traffic accidents. In addition, our dataset provides extensive coverage over 8 states. Other spatiotemporal datasets are mostly constructed for a single city. In contrast, our dataset covers a broader range of areas and facilitates the study of cross-sectional trends beyond a single area.
For details, we describe a comparison with existing spatiotemporal datasets in Table \ref{tab_dataset_st}.

\begin{table}[t!]
\centering
\caption{Comparison between our dataset vs. several existing spatiotemporal datasets.} \label{tab_dataset_st}
\resizebox{\columnwidth}{!}{
\begin{small}
\begin{tabular}{@{}p{1.0 cm} p{2.3 cm} p{2.3 cm} p{2.3 cm} p{2.3 cm} p{2.3 cm}@{}}
\toprule
 & METR-LA \cite{li2017diffusion} & PEMS-Bay \cite{li2017diffusion} &   Taxi NYC \cite{sun2020predicting} & Didi Chuxing \cite{li2018multi} & Our Dataset \\
\midrule
Targets & {Traffic speed and volume} & {Traffic speed and volume} & {\# Rides in a region} & {Trip time} & {Traffic accident} \\
\midrule
Coverage & 6 million for 4 months from Mar. 2012 to Jun. 2012 & 16 million for 6 months: Jan. 2017 - May 2017 & 35 million over 5 years: Jan. 2011 - Jun.  2016 &  61.4 million over six months: May 2017 - Oct. 2017 & 9 million in 8 states up to 20 years until  2023 \\
\midrule
Data Source & Highway loop detectors & CalTrans &  Taxi GPS data for NYC &  Didi
Chuxing Beijing & Department of Transportation \\
\midrule
Features &  Traffic speed and volume & Traffic speed and volume & Traffic flow, latitude, longitude, distances  & Travel distance, \# road, lights, turns for each ride & Traffic volume, weather, and road network features \\
\bottomrule
\end{tabular} 
\end{small}
}
\end{table}

\section{Discussions}

\textbf{Interpretations and future directions.}
It is worth noting that our goal is not to pinpoint exactly where new accidents will occur, as this is clearly not feasible.
Instead, our results can be interpreted as providing suggestions for the risk or hazard of a particular road, whether or not an accident will occur \cite{he2021inferring}.
We can use the predicted accident count as a risk oracle for guiding driving behaviors \cite{balakrishnan2017making}.
A more accurate risk model can also inform decision-making when it comes to insurance assessment.

With this in mind, we discuss several promising questions for future work.
First, one can consider the accident prediction problem with our dataset and methodology, but at different granularities.
One can capture variability within a week by predicting at a daily or even an hourly level \cite{zhou2020riskoracle}.
This can be used to compare risks between weekdays vs. weekends or rush hour vs. evening hours.
Second, one can develop novel time series graph learning methods, such as autoregressive models \cite{ihueze2018road}.
Besides predicting crash frequency, another relevant statistic is the crash rate as normalized by vehicle mileage. %
Third, it would be interesting to explore whether recent developments in road network construction, as captured by satellite images, can inform the design of new predictive features.
More broadly, there are ample opportunities to use our datasets to study policy interventions on road safety.

\smallskip
\textbf{The ML4RoadSafety package.} To facilitate further research, we have developed a package to make our datasets easily accessible to researchers.  
Our package utilizes the same data format as the existing graph learning library, {PyTorch Geometric}, ensuring full compatibility with {PyTorch}.
A user can access our dataset with a single line of code by specifying the name of a state, such as Massachusetts. The package will automatically download, store, and return a dataset object. 
Our package then offers functions to obtain accident records and network features for a particular month.
Besides the datasets, we provide a trainer module to train and evaluate a graph neural network on our datasets. We also provide the implementations of multiple existing graph neural networks. Using the trainer, the user can easily implement other training techniques, including multitask learning, transfer learning, and contrastive learning. We provide code examples in Appendix \ref{sec_use_package}.

\section{Conclusion}

We collect a large-scale dataset of 9 million traffic accident records across eight states to analyze traffic accident occurrences using road networks, traffic flow, and weather reports. Through extensive experiments, we found that existing graph neural networks can be used to predict accident labels with an over 87\% AUROC score. This uses multitask and transfer learning techniques on graphs. Our analysis reveals strong cross-sectional similarities across states regarding road network structures. Ablation studies validate the importance of graph-structural features for achieving the results. %

\paragraph{Acknowledgement.} This work has been inspired by a distinguished seminar talk delivered by Prof. Hari Balakrishnan at Northeastern University, who described the problem concerning road safety \cite{balakrishnan2017making}. H. Z. would like to thank Prof. Ravi Sundaram for several discussions during various steps of this project. This research is supported in part by Northeastern University's Transforming Interdisciplinary Experiential Research (TIER) 1: Seed Grant/Proof of Concept Program.

\begin{refcontext}[sorting=nyt]
\printbibliography
\end{refcontext}
\appendix
\section{Dataset Collection Procedure}\label{sec_data_construction}

In this section, we describe the details of our dataset collection process.
We have provided the implementation in our code repository: \url{https://github.com/VirtuosoResearch/ML4RoadSafety}.
We have also uploaded the entire dataset to an open repository:
\url{https://doi.org/10.7910/DVN/V71K5R}.
This section serves as the documentation for the collection process.
For reference, we provide below the links to public data sources used to construct our datasets.

\begin{table}[!h]
\centering
\caption{Links to the data sources to construct our datasets.}\label{table_data_source}
\resizebox{0.99\textwidth}{!}
{
\begin{tabular}{l p{10.0cm}}
\toprule
\toprule
Traffic accident records \\
\midrule
Delaware Open Data & \url{https://data.delaware.gov/Transportation/Public-Crash-Data-Map/3rrv-8pfj}\\
Delaware DOT & \url{https://deldot.gov/search/}\\
Iowa DOT & \url{https://icat.iowadot.gov/}\\
Illinois DOT & \url{https://gis-idot.opendata.arcgis.com/search?collection=Dataset&q=Crashes}\\
Mass DOT & \url{https://apps.impact.dot.state.ma.us/cdp/home}\\
Maryland Open Data & \url{https://opendata.maryland.gov/Public-Safety/Maryland-Statewide-Vehicle-Crashes/65du-s3qu}\\
MN Crash & \url{https://mncrash.state.mn.us/Pages/AdHocSearch.aspx}\\
Montana DOT & \url{https://www.mdt.mt.gov/publications/datastats/crashdata.aspx}\\
Nevada DOT & \url{https://ndot.maps.arcgis.com/apps/webappviewer/index.html?id=00d23dc547eb4382bef9beabe07eaefd}\\
\midrule
\midrule
Road networks \\
\midrule
OSMnx Street Network Dataverse & \url{https://dataverse.harvard.edu/dataset.xhtml?persistentId=doi:10.7910/DVN/CUWWYJ} \\
OpenStreetMap Road Categories & \url{https://wiki.openstreetmap.org/wiki/Highways\#Classification} \\
{Google Map API} & \url{https://maps.google.com/} \\
\midrule
\midrule
Weather reports \\
\midrule
Meteostat API & \url{https://meteostat.net/en/}\\
\midrule
\midrule
Traffic volume reports \\
\midrule
Delaware & \url{https://deldot.gov/search/} \\
Maryland & \url{https://data-maryland.opendata.arcgis.com/datasets/mdot-sha-annual-average-daily-traffic-aadt-segments/explore?location=38.833256\%2C-77.269751\%2C8.30&showTable=true} \\
Massachusetts & \url{https://mhd.public.ms2soft.com/tcds/tsearch.asp?loc=Mhd&mod=} \quad \url{https://www.mass.gov/lists/massdot-historical-traffic-volume-data} \\
Nevada & \url{https://geohub-ndot.hub.arcgis.com/datasets/trina-stations/explore?location=38.490765\%2C-116.969086\%2C7.27&showTable=true} \\
\bottomrule
\bottomrule
\end{tabular}
}
\vspace{-0.15in}
\end{table}

\subsection{Traffic accident records}

First, we construct the accident labels in our dataset. 
We collect accident records for each state from the respective Department of Transportation websites. Each accident is associated with a report detailing the date and geographic coordinates (i.e., latitude and longitude) of the accident. 
We collect accident data from eight states. The records are available in the sources listed in Table \ref{table_data_source}.
Then, we map each accident to the nearest road segment in the road network based on its coordinates. 
Specifically, for an accident located at point $c$, and for a road edge defined by its two endpoints $a$ and $b$, we assign the accident to the edge that maximizes $D(a,b) - (D(a,c) + D(b,c))$, where $D(\cdot)$ is the Euclidean distance between two locations.

\subsection{Road networks}

We generate a road network for a state as a directed graph, based on the road network structure in the state extracted from OpenStreetMap. 
Nodes correspond to all the publicly available road intersections, and edges correspond to the road segments between these nodes. Therefore, one road can have multiple edges depending on the number of intersections with other roads. For instance, a road with three intersections would be divided into three edges.

The edges in the graph include road segments in a state, including every city, town, urbanized area, county, census tract, and Zillow neighborhood.
We obtain the above information from the OSMnx Street Networks Dataverse in OpenStreetMap (See Table \ref{table_data_source}).

\subsection{Road network features}

We describe four types of features associated with road networks as follows.

\textbf{Graph-structural features (node-level, static).}
All nodes in a graph are associated with static structural features, including node in-degrees, out-degrees, and betweenness centrality scores. 
Node degree values are encoded as one-hot vectors, with the vector dimension equal to the maximum degree in the graph. Two such vectors are generated for each node to represent its in-degree and out-degree, respectively. Betweenness centrality is represented as a continuous feature, measuring the proportion of all shortest paths between node pairs that pass through a given node.

\textbf{Historical weather information (node-level, temporal).} Besides, each node is also associated with daily weather information. 
We collect the following six features related to the weather conditions within a particular month: (i-iii) the maximum, minimum, and average of the temperature on the road's surface; (iv) the total precipitation, including rainfall or snowfall; (v) the average wind speed; (vi) the sea level air pressure.
For every node, these weather conditions are obtained from the nearest meteorological station based on its geographic coordinates. 
The weather information is extracted using the Meteostat API (cf. Table \ref{table_data_source}).
Given the temporal nature of this information and its availability for every node, we consider these weather features to be particularly important for our predictive task. %

\textbf{Road information (edge-level, static).}
Each edge is associated with static features that describe its physical length and road category.  
The length of a road segment is represented as a real-valued feature measured in meters. Road categories are encoded as a $24$-dimensional binary vector, where each entry indicates whether the edge belongs to a specific category. Each road segment may belong to one or more of the following $24$ categories: one-way, primary, primary link, secondary, secondary link, access ramp, bus stop, crossroad, disused, elevator, escape, living street, motorway, motorway link, residential, rest area, road, stairs, tertiary, tertiary link, trunk, trunk link, unsurfaced, and unclassified. The road information is obtained from OpenStreetMap.

\textbf{Traffic volume (edge-level, temporal).} Each edge is also associated with a traffic volume feature measured on an annual basis. This feature represents the average number of vehicles traveling along the corresponding road segment per year and is encoded as a real-valued attribute. The traffic volume data are obtained from the Annual Average Daily Traffic (AADT) reports published by the Department of Transportation of each state. These reports provide traffic counts for a subset of streets within the state. Using the Google Maps API, we retrieve the geographic coordinates corresponding to each street name and map them to the edges in our road network.

\subsection{Summary of road network features}
Here is a list of node-level features we have included in our dataset:
latitude,
longitude,
node indegree and outdegree,
betweenness centrality,
average surface temperature,
max surface temperature,
min surface temperature,
total precipitation,
average wind speed, and
sea level air pressure.

Here are the edge-level features in our dataset:
a binary label that indicates whether the road is one-way or not,
a multi-class label that indicates the road category,
length of the road, and
annual average daily traffic (AADT) --- if this information is available in the report.

\section{Experiment Details} \label{sec_detailed_exp}

In this section, we describe the details of our experiments that were left out in the main text. First, we describe additional implementation details. Then, we describe the omitted experimental results. These include evaluations of accident prediction performance using precision and recall scores, detailed results of hyperparameter tuning, preliminary results of contrastive learning, and observations of seasonal variations in accident counts. 
Lastly, we provide code examples of using our package.

\subsection{Implementation details}

In our implementation, we set the dimension of node embeddings as 128, the number of layers as 2, and the hidden dimensionality as 256.  We train our models using Adam as the optimizer. We use a learning rate of 0.001 for 100 epochs on all models. 
The hyperparameters are determined by searching in the following ranges: The hidden dimensionality is tuned in a range of $\set{128, 256, 512}$. The number of layers is tuned in a range of $\set{2, 3, 4}$. The learning rate is tuned in a range of $\set{0.01, 0.001, 0.0001}$.

For each state, we evenly split the available period of accidents into training, validation, and test sets. Table \ref{tab_data_splitting} summarizes the dataset splitting for each state.  While our evaluation focuses on monthly predictions, our datasets can also be utilized for conducting analyses at daily or annual levels.

\begin{table}[h!]
\centering
\caption{Data splitting of accident records of eight states.} \label{tab_data_splitting}
\begin{small}
\begin{tabular}{@{}lcccccccc@{}}
\toprule
& Train (years) & Train (records) & Valid (years) & Valid (records) & Test (years) & Test (records) \\
\midrule
DE & 2009 - 2012 & 112,670 & 2013 - 2017 & 174,278 & 2018 - 2022 & 171,311 \\
IA & 2013 - 2016 & 213,019 & 2017 - 2019 & 171,455 & 2020 - 2022 & 156,065\\
IL & 2012 - 2014 & 856,057 & 2015 - 2017 & 949,745 & 2018 - 2021 & 1,174,900\\
MA & 2002 - 2008 & 1,265,895 & 2009 - 2015 & 933,786 & 2016 - 2022 & 1,096,885 \\
MD & 2015 - 2017 & 341,902 & 2018 - 2019 & 229,446 & 2020 - 2022 & 306,995\\
MN & 2015 - 2017 & 148,361 & 2018 - 2019 & 154,150 & 2020 - 2022 & 188,858 \\
MT & 2016 - 2017 &  40,040 & 2018 &  20,677 & 2019 - 2020 & 39,222 \\
NV & 2016 - 2017 & 101,975 & 2018 & 48,854 & 2019 - 2020 & 86,509 \\
\bottomrule
\end{tabular}
\end{small}
\end{table}

\textbf{Number of parameters in each model.} %
We summarize the model names and their corresponding numbers of parameters as follows. 
MLP, Node2Vec, DeepWalk: $75\times 10^3$;
GCN, GraphSAGE: $253\times 10^3$;
GIN: $778\times 10^3$;
{AGCRN}: $294\times 10^3$;
{STGCN}: $713\times 10^3$;
{Graph Wavenet}: $1,154\times 10^3$;
DCRNN: $1,463\times 10^3$.

\subsection{Additional experimental results}

\begin{table}[t!]
\centering
\caption{We report the precision and recall scores on the test split on eight states, using node embedding methods, graph neural networks, and multitask and transfer learning methods. Each experiment is conducted with three different random seeds. We report the average results along with their standard deviations.} \label{tab_accident_precision_recall}
\resizebox{\columnwidth}{!}{ 
\begin{tabular}{@{}lcccccccc@{}}
\toprule
Precision & DE & IA & IL & MA & MD & MN & MT & NV \\
Training Size & 93,184 & 187,046 & 646,739 & 540,682 & 283,226 & 124,435 & 34,475 & 73,164\\
Positive Rate & 0.23 & 0.07 & 0.14 & 0.10 & 0.15 & 0.05 & 0.05 &  0.12 \\
\midrule
MLP & 4.99$\pm$0.1 & 1.78$\pm$0.0 & 2.54$\pm$0.2 & 3.52$\pm$0.6 & 3.47$\pm$0.1 & 1.87$\pm$0.0 & 0.87$\pm$0.0 & 3.30$\pm$0.0\\
Node2Vec &  12.76$\pm$0.2 & 3.16$\pm$0.3 & 3.52$\pm$0.5 & 3.28$\pm$0.8 & 5.64$\pm$0.4 & 2.38$\pm$0.3 & 3.50$\pm$0.2 & 10.74$\pm$0.0\\
DeepWalk & 14.13$\pm$0.5 & 2.90$\pm$0.5 & 3.37$\pm$0.6 & 3.30$\pm$0.1 & 4.88$\pm$0.1 & 2.64$\pm$0.3 & 2.62$\pm$0.4 & 7.74$\pm$0.0\\
GCN & 11.09$\pm$0.7 & 2.36$\pm$0.1 & 7.54$\pm$0.7 & 4.05$\pm$0.4 & 5.60$\pm$0.1 & 4.60$\pm$0.1 & 5.27$\pm$0.2 & 9.17$\pm$0.1\\
GraphSAGE & 18.56$\pm$0.9 & 4.13$\pm$0.2 & 8.54$\pm$0.3 & 4.71$\pm$0.2 & 7.04$\pm$0.1 & 6.26$\pm$0.4 & 4.11$\pm$0.5 & 8.62$\pm$0.0 \\
GIN &  13.95$\pm$0.6 & 3.63$\pm$0.2 & 9.84$\pm$0.8 & 4.45$\pm$0.8 & 6.50$\pm$0.5 & 4.08$\pm$0.7 & 5.72$\pm$0.7 & 10.27$\pm$0.0 \\
AGCRN & 11.36$\pm$0.6 & 3.50$\pm$0.2 & 7.60$\pm$1.1 & 4.30$\pm$0.3 & 6.61$\pm$0.2 & 5.66$\pm$0.2 & 3.12$\pm$0.1 & 7.74$\pm$0.1 \\
STGCN & 12.33$\pm$0.2 & 5.20$\pm$0.2 & 7.14$\pm$1.7 & 4.54$\pm$0.9 & 8.91$\pm$0.1 & 4.65$\pm$0.1 & 4.04$\pm$0.9 & 9.72$\pm$0.9 \\
Graph Wavenet & 15.56$\pm$0.5 & 3.83$\pm$0.3 & 7.76$\pm$0.7 & 4.22$\pm$1.6 & 7.21$\pm$0.3 & 6.62$\pm$0.5 & 3.36$\pm$0.1 & 7.84$\pm$0.1 \\ 
DCRNN &  11.33$\pm$0.5 & 3.87$\pm$0.4 & 6.16$\pm$0.1 & 4.67$\pm$0.1 & 4.75$\pm$0.4 & 3.62$\pm$0.1 & 3.66$\pm$0.4 & 10.06$\pm$0.1 \\
\midrule \midrule
STL w/ GraphSAGE & 18.56$\pm$0.9 & 4.13$\pm$0.2 & 8.54$\pm$0.3 & 4.71$\pm$0.2 & 7.04$\pm$0.1 & 6.26$\pm$0.4 & 4.11$\pm$0.5 & 8.62$\pm$0.0 \\
MTL w/ GraphSAGE  &   14.34$\pm$0.1 & 3.52$\pm$0.4 & \textbf{13.62$\pm$0.7} & {5.11$\pm$0.1} & {8.66$\pm$0.0} & 4.44$\pm$0.2 & \textbf{5.85$\pm$0.3} & {16.80$\pm$0.0} \\
MTL-FT w/ GraphSAGE & \textbf{18.61$\pm$0.1} & \textbf{4.66$\pm$0.0} & 13.61$\pm$0.5 & \textbf{5.20$\pm$0.1} & \textbf{8.76$\pm$0.1} & \textbf{6.66$\pm$0.3} & 5.72$\pm$0.0 & \textbf{16.85$\pm$0.4}\\
TL w/ GraphSAGE &  14.10$\pm$0.4 & - & - & 5.07$\pm$0.4 & 8.51$\pm$0.4 & - & - & 9.5$\pm$0.1\\
\midrule\midrule
Recall  & DE & IA & IL & MA & MD & MN & MT & NV \\
\midrule
MLP &  60.4$\pm$3.7 & 83.8$\pm$2.8 & 81.1$\pm$2.5 & 79.9$\pm$1.4 & 67.2$\pm$2.0 & 70.3$\pm$1.2 & 72.0$\pm$1.6 & 74.3$\pm$0.4\\
Node2Vec &  \textbf{83.8$\pm$1.3} & \textbf{89.5$\pm$1.6} & 78.2$\pm$0.6 & 80.1$\pm$3.1 & 80.0$\pm$2.3 & \textbf{80.0$\pm$2.2} & 73.6$\pm$0.9 & 83.9$\pm$0.5  \\
DeepWalk & 83.2$\pm$2.9 & 80.7$\pm$4.8 & 81.0$\pm$1.1 & 85.7$\pm$3.1 & \textbf{86.2$\pm$3.6} & 78.4$\pm$2.9 & 74.0$\pm$3.3 & \textbf{88.9$\pm$0.0}  \\
GCN &  75.2$\pm$3.0 & 74.0$\pm$2.0 & \textbf{84.1$\pm$0.8} & 82.5$\pm$2.1 & 79.1$\pm$2.6 & 70.7$\pm$1.8 & 63.6$\pm$3.8 & 85.4$\pm$0.8 \\ 
GraphSAGE & 60.9$\pm$2.7 & 58.5$\pm$2.0 & 72.7$\pm$2.4 & 51.0$\pm$1.3 & 68.7$\pm$1.1 & 54.5$\pm$2.2 & 59.6$\pm$2.9 & 84.7$\pm$1.1  \\
GIN & 65.8$\pm$3.7 & 75.0$\pm$2.4 & 78.0$\pm$4.3 & 48.2$\pm$2.0 & 78.3$\pm$3.2 & 74.8$\pm$2.4 & 65.6$\pm$1.8 & 88.6$\pm$0.9 \\
AGCRN & 71.5$\pm$1.6 & 71.7$\pm$1.9 & 73.4$\pm$1.0 & 58.2$\pm$1.6 & 75.4$\pm$1.5 & 73.1$\pm$1.5 & 70.7$\pm$1.3 & 82.6$\pm$3.1 \\
STGCN & 82.9$\pm$0.2 & 78.2$\pm$4.7 & 75.3$\pm$1.2 & 60.7$\pm$1.3 & 77.6$\pm$1.0 & 75.8$\pm$0.3 & 72.6$\pm$1.1 & 83.0$\pm$0.0\\
Graph Wavenet  & 73.5$\pm$3.0 & 67.7$\pm$0.7 & 78.3$\pm$0.2 & 77.7$\pm$1.1 & 73.0$\pm$0.2 & 72.9$\pm$3.2 & 66.1$\pm$1.0 & 79.3$\pm$0.1 \\
DCRNN &  75.4$\pm$2.6 & 71.1$\pm$2.2 & 80.5$\pm$1.7 & 81.5$\pm$2.0 & 83.0$\pm$1.4 & 63.6$\pm$1.3 & 66.3$\pm$1.0 & 83.4$\pm$0.8 \\
\midrule \midrule
STL w/ GraphSAGE & 60.9$\pm$2.7 & 58.5$\pm$2.0 & 72.7$\pm$2.4 & 51.0$\pm$1.3 & 68.7$\pm$1.1 & 54.5$\pm$2.2 & 59.6$\pm$2.9 & 84.7$\pm$1.1  \\
MTL w/ GraphSAGE & 63.8$\pm$1.8 & 78.6$\pm$1.0 & 75.6$\pm$1.3 & 75.5$\pm$0.9 & 78.1$\pm$1.2 & 77.6$\pm$0.7 & 74.2$\pm$0.9 & 82.3$\pm$0.5\\
MTL-FT w/ GraphSAGE & 64.5$\pm$0.7 & 78.2$\pm$2.5 & 76.7$\pm$1.2 & 	73.7$\pm$1.1 & 77.4$\pm$2.1 & 73.2$\pm$0.8 & \textbf{74.3$\pm$3.6} & 84.9$\pm$1.2\\
TL w/ GraphSAGE & 66.7$\pm$1.4 & - & - & \textbf{92.7$\pm$1.3} & 81.7$\pm$1.6 & - & - & 84.1$\pm$3.3\\
\bottomrule
\end{tabular} 
}
\end{table}

\begin{table}[t!]
\centering
\caption{Ablation study of different hyperparameters, including the number of layers, the hidden dimensionality, the learning rate, and training epochs. We report the AUROC scores on the validation split on the Delaware state dataset using GraphSAGE and DCRNN. Each experiment is conducted with three different random seeds. We report the average results along with their standard deviations.
} \label{tab_hyperparams}
\begin{small}
\begin{tabular}{@{}lccc|cccccc@{}}
\toprule
& \multicolumn{3}{c|}{GraphSAGE} & \multicolumn{3}{c}{DCRNN}  \\ \midrule
Number of layers & $2$ & $3$ & $4$ & $2$ & $3$ & $4$ \\ \midrule 
& \textbf{85.2}$\pm$0.1 & 84.9$\pm$0.3 & 84.4$\pm$0.4 & \textbf{67.8}$\pm$1.2 & 67.2$\pm$0.8 & 67.3$\pm$0.5 \\ \midrule \midrule
Hidden dimensionality & $128$ & $256$ & $512$ & $128$ & $256$ & $512$  \\ \midrule
& 84.5$\pm$0.4 & \textbf{85.2}$\pm$0.1 & 84.5$\pm$0.5 & 66.9$\pm$0.7 & \textbf{67.8}$\pm$1.2 & 66.9$\pm$1.1\\ \midrule\midrule
Learning rate & $1e^{-2}$ & $1e^{-3}$ & $1e^{-4}$ & $1e^{-2}$ & $1e^{-3}$ & $1e^{-4}$ \\ \midrule
& 85.0$\pm$0.7 & \textbf{85.2}$\pm$0.1 & 84.0$\pm$0.5 & 66.8$\pm$1.0 & \textbf{67.8}$\pm$1.2 & 66.5$\pm$0.9 \\ \midrule\midrule
Epochs & $50$ &  $100$ & $200$ & $50$ &  $100$ & $200$ \\ \midrule
& 84.0$\pm$0.2 & \textbf{85.2}$\pm$0.1 & \textbf{85.2}$\pm$0.3 & 66.4$\pm$0.7 & \textbf{67.8}$\pm$1.2 & \textbf{67.8}$\pm$1.0 \\ 
\bottomrule
\end{tabular} 
\end{small}
\end{table}

\smallskip
\textbf{Detailed results of classification metrics.} Next, we report the detailed results of classifying whether an accident occurred on a particular road. Table \ref{tab_accident_precision_recall} reports the recall and precision scores of the predictions. First, we observe that for all baselines, the recall scores are higher than the precision scores. Using the graph neural networks can predict whether an accident occurred on a road with 10\% precision and 85\% recall on average over eight states. 
Second, we also observe that multitask learning outperforms training a single model per state by 20\% and 21\% in terms of precision and recall, respectively.
Third, combining traffic volume with accident prediction also improves training a model only for traffic prediction by 4\% and 15\% in terms of precision and recall scores. 

We observe that while using MLP on node embeddings achieves higher recall than graph neural networks, graph neural networks achieve higher precision scores. This indicates that MLP models tend to be more overconfident when classifying the likelihood of an accident occurring on a road. Given the low precision score, we report the AUROC score in the main text, which provides a summary of the recall score across all decision thresholds.

\textbf{Detailed results of hyperparameter tuning.}
We conduct an ablation study on the hyperparameters used in our experiments, including the number of layers, the hidden dimensionality, the learning rate, and the number of epochs. In each ablation study, we vary one hyperparameter and keep the others unchanged. The fixed hyperparameters are used as follows: the number of layers of 2, the hidden dimensionality of 256, a learning rate of $1e^{-3}$, and 100 epochs.

Table \ref{tab_hyperparams} shows the validation AUROC scores, varying hyperparameters for GraphSAGE and DCRNN on the Delaware state dataset. We notice that using the number of layers as 2, hidden dimensionality as 256, and learning rate as $1e^{-3}$ yields the best results for both baselines. The validation performance stops improving after training the model up to 100 epochs. We also find that these hyperparameters are useful for other models. Thus, we adopt these settings as the default parameters in the experiments.

\smallskip
\textbf{Applying graph contrastive learning.} We conduct a preliminary study of applying graph contrastive learning using GraphSAGE and DCRNN as the base model across four states. We compare them with the supervised learning baselines. Table \ref{tab_constrastive_learning} shows the results.
We find that graph contrastive learning can improve the test performance of the baselines in some states. 
\begin{table}[t!]
\centering
\caption{We report the results of applying graph contrastive learning on our datasets using GraphSAGE and DCRNN. 
We report the AUROC scores on the test split across four states. Each experiment is conducted with three different random seeds. We report the average results along with their standard deviations.}\label{tab_constrastive_learning}
\begin{small}
\begin{tabular}{@{}lcccccccc@{}}
\toprule
 & DE & MA & MD & NV \\ \midrule
GraphSAGE & 87.6$\pm$0.1 & 81.8$\pm$0.1 & 87.5$\pm$0.0 & 91.6$\pm$0.9 \\
GraphSAGE w/ GCL & 86.7$\pm$0.2 & 82.4$\pm$0.8 & 85.9$\pm$0.4 & 91.8$\pm$0.4 \\ \midrule
DCRNN & 81.2$\pm$1.2  & 70.5$\pm$0.1 & 84.5$\pm$0.3 & 90.5$\pm$0.7\\
DCRNN w/ GCL & 86.6$\pm$0.3 & 82.5$\pm$0.7 & 87.8$\pm$0.9 & 91.7$\pm$0.4 \\ \midrule
STGCN &  85.4$\pm$0.1 & 81.9$\pm$0.3 & 88.7$\pm$0.1 & 91.5$\pm$0.3\\ 
STGCN w/ GCL & 86.0$\pm$0.1 & 81.7$\pm$0.1 & 89.7$\pm$0.5 & 92.4$\pm$0.1\\ 
\bottomrule
\end{tabular} 
\end{small}
\end{table}

\begin{figure}[h!]
     \centering
     \begin{subfigure}[b]{0.24\textwidth}
         \centering
         \includegraphics[width=\textwidth]{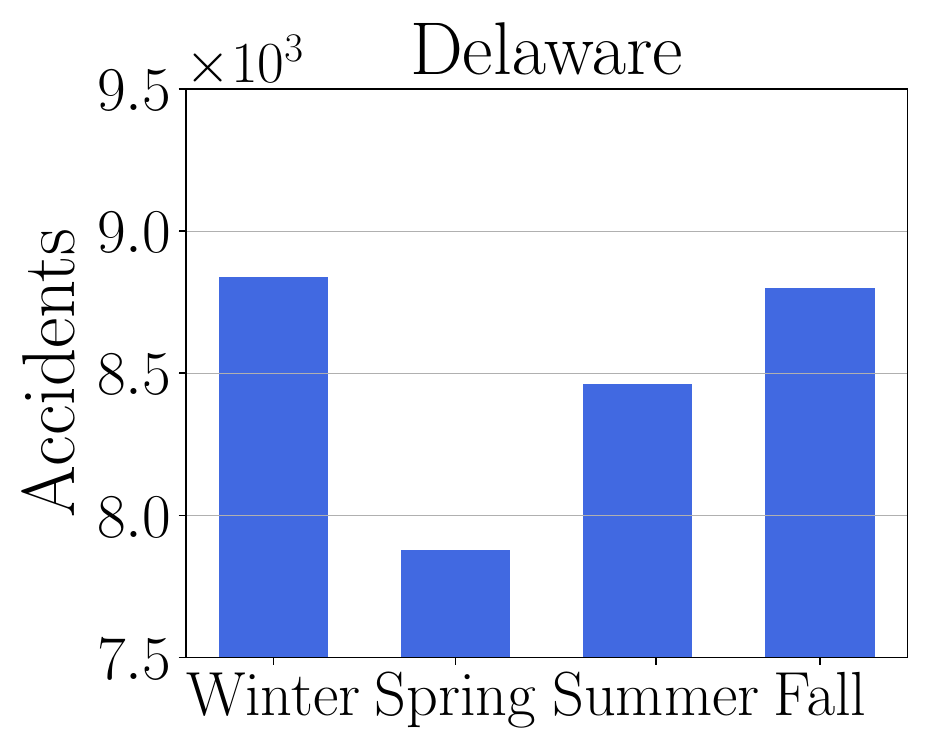}
     \end{subfigure}
     \hfill
     \begin{subfigure}[b]{0.24\textwidth}
         \centering
         \includegraphics[width=\textwidth]{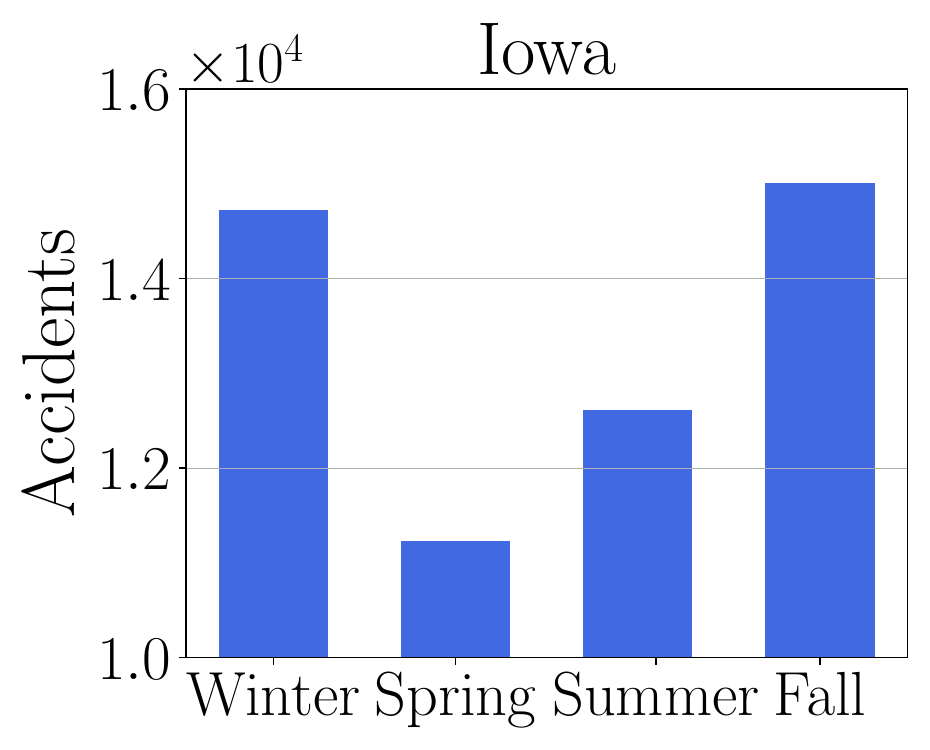}
     \end{subfigure}
     \hfill
     \begin{subfigure}[b]{0.24\textwidth}
         \centering
         \includegraphics[width=\textwidth]{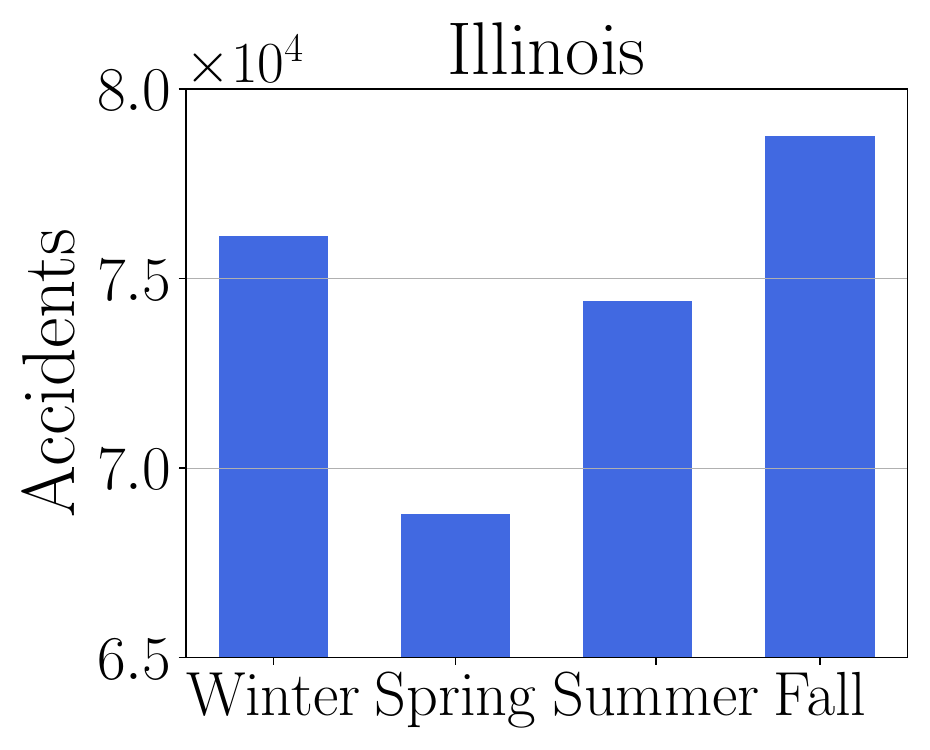}
     \end{subfigure}
     \hfill
     \begin{subfigure}[b]{0.24\textwidth}
         \centering
         \includegraphics[width=\textwidth]{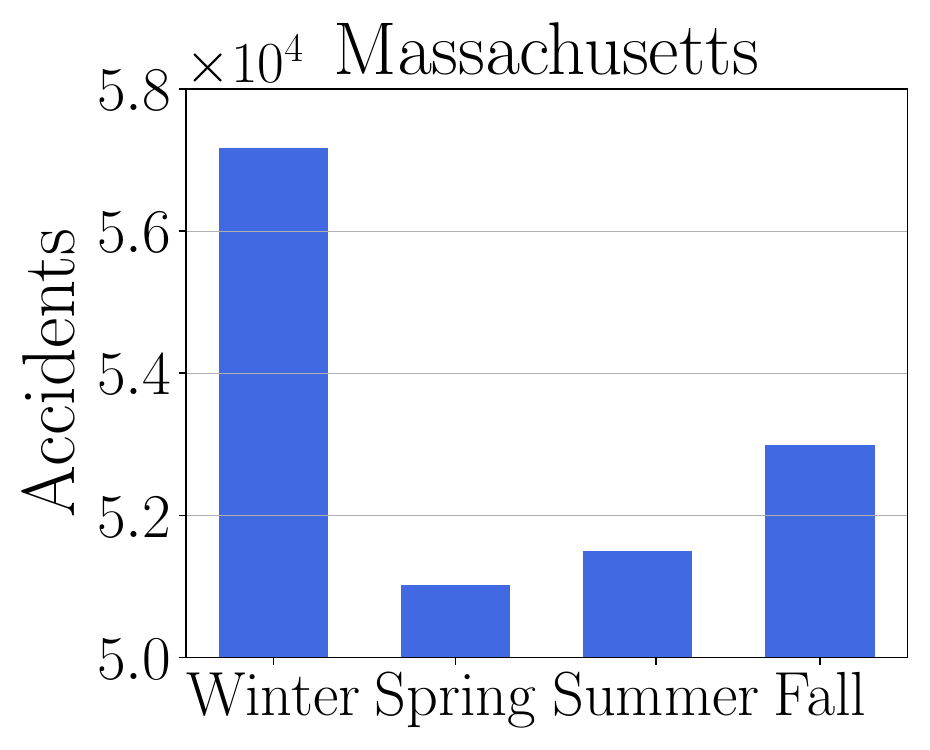}
     \end{subfigure}
     \begin{subfigure}[b]{0.24\textwidth}
         \centering
         \includegraphics[width=\textwidth]{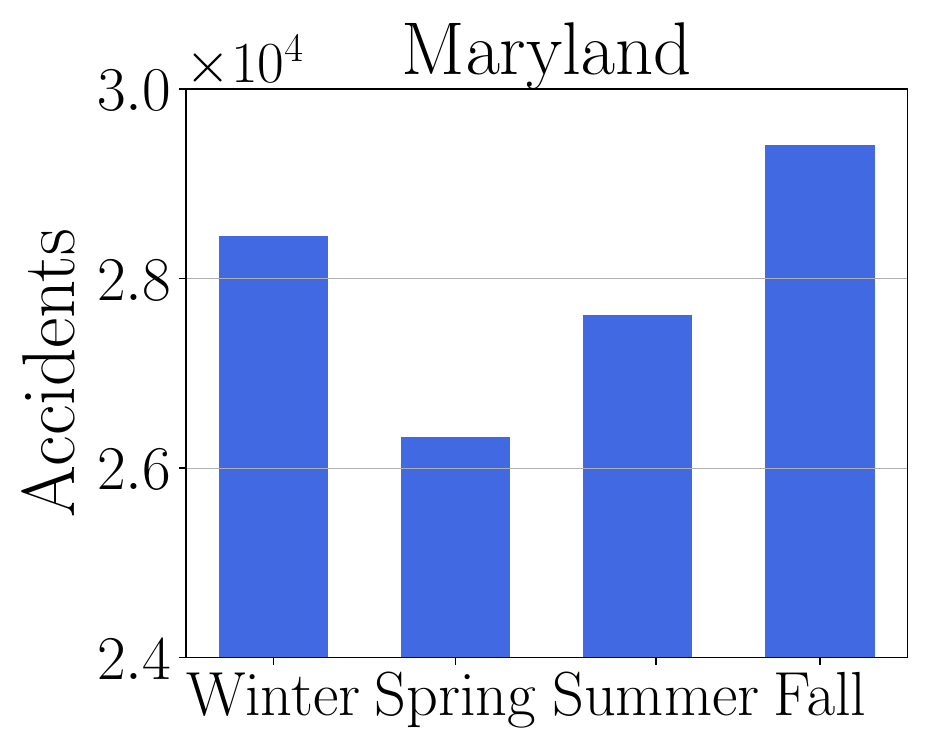}
     \end{subfigure}
     \hfill
     \begin{subfigure}[b]{0.24\textwidth}
         \centering
         \includegraphics[width=\textwidth]{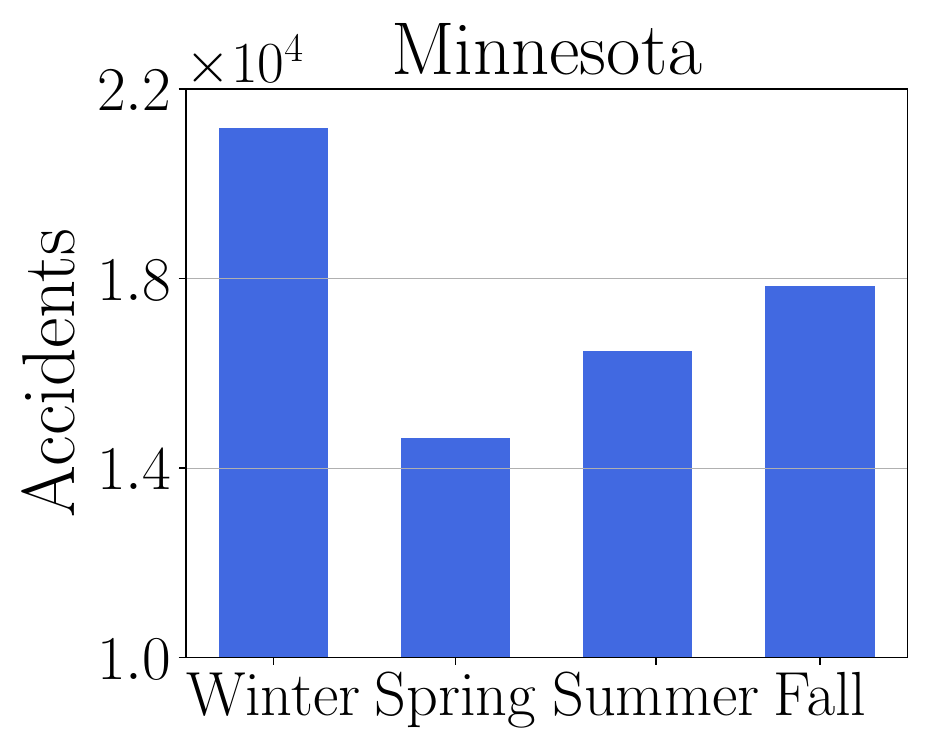}
     \end{subfigure}
     \hfill
     \begin{subfigure}[b]{0.24\textwidth}
         \centering
         \includegraphics[width=\textwidth]{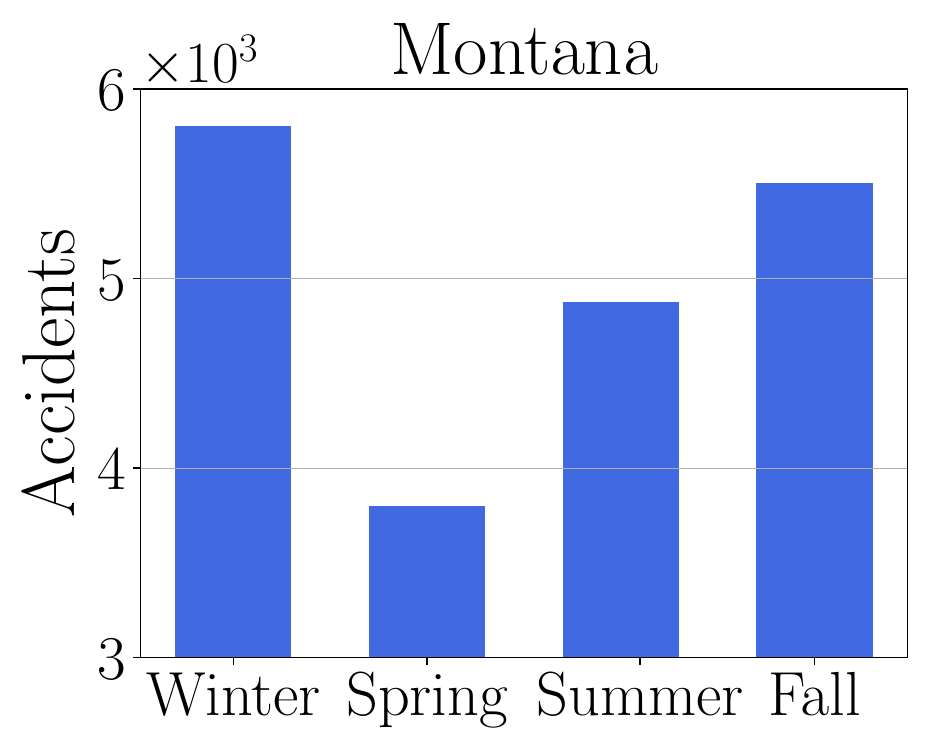}
     \end{subfigure}
     \hfill
     \begin{subfigure}[b]{0.24\textwidth}
         \centering
         \includegraphics[width=\textwidth]{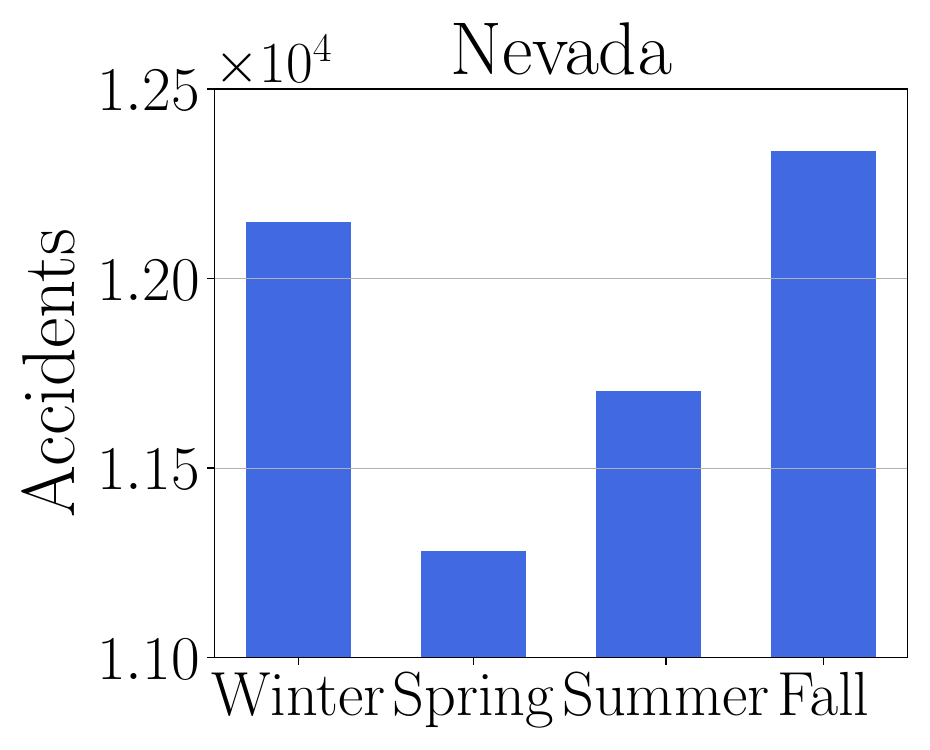}
     \end{subfigure}
     \caption{We illustrate the seasonal trend of accident counts within a year. Across all eight states, we consistently observe higher accident counts during Winter and Fall compared to Spring and Summer.}\label{fig_monthly_accident_trend}     
\end{figure}

\smallskip
\textbf{Seasonal patterns of road accidents.} 
We study how the number of accidents evolves within a year and explore its potential association with seasonal patterns. 
We hypothesize that the accidents may be affected by the weather and show seasonal trends. 
To examine this, we aggregate accident counts within four seasons in a year for each state. Specifically, we aggregate accidents occurring from December to February for winter, from March to May for spring, from June to August for summer, and from September to November for fall. 
Figure \ref{fig_monthly_accident_trend} shows trends of accident numbers across the seasons for the four states. 
We notice a significant disparity in accident counts between winter and fall compared to spring and summer. 
The disparity suggests that accidents may indeed be influenced by seasonal factors, including severe weather conditions and road hazards that are more prevalent during the colder months.

\subsection{Code examples}\label{sec_use_package}

We provide examples for accessing the data and training models using our dataset. 
Our package uses the same data format as the existing graph learning library, {PyTorch Geometric}, which is fully compatible with {PyTorch}.
As shown in Code Snippet \ref{code}, users can access a dataset with a single line of code by specifying the name of a state, such as Massachusetts.
The package will automatically download, process, and return the dataset object. The dataset objective provides respective access to the road network graph structure, road network features, and the corresponding accident labels.
In addition, our package provides a function within the dataset object to obtain the accident records and network features for a particular month.

\medskip

\begin{lstlisting}[language=Python, caption=ML4RoadSafety data loader, label=code]
>>> from ml_for_road_safety import TrafficAccidentDataset
# Creating the dataset as a PyTorch Geometric dataset object
>>> dataset = TrafficAccidentDataset(state_name = "MA")
# Loading the accident records and traffic network features of a particular month
>>> data = dataset.load_monthly_data(year = 2022, month = 1)
# Pytorch Tensors storing the list of edges with accidents and accident numbers
>>> accidents, accident_counts = data["accidents"], data["accident_counts"]
# Pytorch Tensors of node features, edge list, and edge features
>>> x, edge_index, edge_attr = data["x"], data["edge_index"], data["edge_attr"]
\end{lstlisting}

Our package includes a trainer class designed to train graph neural networks on data from a single state in our dataset. This class encapsulates the full training and evaluation pipeline. As illustrated in Code Snippet~\ref{code_trainer}, users can instantiate a trainer object by specifying a model, a dataset, and an evaluation metric. The training process can then be initiated with a single function call, which automatically executes the training and returns the evaluation results upon completion.
\begin{lstlisting}[language=Python, caption=ML4RoadSafety trainer, label=code_trainer]
>>> from ml_for_road_safety import Trainer, Evaluator, TrafficAccidentDataset
# Creating the dataset 
>>> dataset = TrafficAccidentDataset(state_name = "MA")
# Get an evaluator for accident prediction, e.g., the classification task. 
>>> evaluator = Evaluator(type = "classification")
# Initialize a trainer with a GNN model, a dataset, and an evaluator
>>> trainer = Trainer(model, dataset, evaluator, ...)
# Conduct training and evaluation inside the trainer
>>> log = trainer.train()
\end{lstlisting}

\subsubsection{Multitask and transfer learning} 
In multitask learning, we combine multiple datasets and optimize the average loss computed across all tasks. Our implementation proceeds as follows: a separate trainer is instantiated for each dataset, and the average loss is optimized by iterating through all task trainers within each epoch.
To make this easy to use, we have integrated the logic into a multitask learning trainer. As shown in Code Snippet \ref{code_multitask_trainer}, users can create a multitask trainer by specifying a model and providing a list of datasets. The multitask model can then be trained with a single function call.

\begin{lstlisting}[language=Python, caption=Implementation of multitask learning, label=code_multitask_trainer]
>>> from ml_for_road_safety import MultitaskTrainer
# Specify the tasks that are combined in multitask learning
>>> task_list = ["MA_accident_classification", "MD_accident_classification", ...]
>>> task_datasets = {}; task_evaluators = {}
>>> for task_name in task_list: 
>>>     state_name, data_type, task_type = task_name.split("_")
>>>     task_datasets[task_name] =  TrafficAccidentDataset(state_name = state_name)
>>>     task_evaluators[task_name] = Evaluator(type=task_type)
# Initialize a trainer with a GNN model, multiple datasets, and multiple evaluators
>>> trainer = MultitaskTrainer(model, tasks = task_list, task_to_datasets=task_datasets, task_to_evaluators=task_evaluators, ...) 
# Conduct multitask learning and evaluation for every task
>>> trainer.train()
\end{lstlisting}

We apply transfer learning to leverage knowledge from traffic volume prediction for improving traffic accident prediction.
We implement this by training a single model on both tasks simultaneously. As shown in Code Snippet \ref{code_transfer}, users can create a multitask learning trainer for a single state that includes both accident prediction and traffic volume prediction. The trainer then optimizes the model by minimizing the average loss across the two tasks.
\begin{lstlisting}[language=Python, caption=Implementation of transfer learning, label=code_transfer]
>>> from ml_for_road_safety import MultitaskTrainer 
# Specify the accident and volume prediction tasks from one state
>>> task_list = ["MA_accident_classification", "MA_volume_regression"]
>>> task_datasets = {}; task_evaluators = {}
>>> for task_name in task_list: 
>>>     state_name, data_type, task_type = task_name.split("_")
>>>     task_datasets[task_name] =  TrafficAccidentDataset(state_name = state_name)
>>>     task_evaluators[task_name] = Evaluator(type=task_type)
# Initialize a trainer with a GNN model and two tasks of accident and volume prediction. 
>>> trainer = MultitaskTrainer(model, tasks = task_list, task_to_datasets=task_datasets, task_to_evaluators=task_evaluators, ...)
# Conduct multitask learning and evaluation for both tasks
>>> trainer.train()
\end{lstlisting}

\subsubsection{Graph contrastive learning}

Our package can be easily extended to incorporate graph contrastive learning methods for traffic accident prediction using our datasets. 
For example, graph contrastive learning \cite{you2020graph} can be implemented with only a few lines of code. As shown in Code Snippet \ref{code_contrastive_learning}, one can define a trainer for contrastive learning by modifying the training loss in the base trainer class. Then, the user can use the trainer to perform contrastive learning on our dataset.
\begin{lstlisting}[language=Python, caption=Implementation of graph contrastive learning, label=code_contrastive_learning]
>>> from ml_for_road_safety import Trainer
# Define a trainer for contrastive learning inherited from the base Trainer class
>>> class GraphContrastiveLearningTrainer(Trainer):
# Modify the training loss in the training logic
>>>     def train_epoch(self):
#           Define the contrastive loss
>>>         ...
>>>         loss = info_nce(outputs_1, outputs_2)
>>>         ...
# Initialize a contrastive learning trainer 
>>> trainer = GraphContrastiveLearningTrainer(model, dataset, evaluator, ...)
# Conduct training and evaluation inside the trainer
>>> log = trainer.train()
\end{lstlisting}

\subsubsection{Spatiotemporal graph neural networks}

Next, our package also supports the evaluation of spatiotemporal graph neural networks. For example, as shown in Code Snippet \ref{code_spatiotemporal}, the user can create a spatiotemporal model using our package, such as STGCN, by specifying the corresponding model name. The package integrates implementations of spatiotemporal models from the open-source repository pytorch-geometric-temporal. Once the model is defined, users can instantiate a trainer object to directly train it on the selected dataset.

\begin{lstlisting}[language=Python, caption=Training a spatiotemporal model, label=code_spatiotemporal]
>>> from ml_for_road_safety import Trainer, GNN
# Create a spatiotemporal model, e.g., STGCN, from an online implementation
>>> model = GNN(encoder = "stgcn", ...) 
# Initialize a trainer with the model and specify use_time_series as True
>>> trainer = Trainer(model, dataset, evaluator, use_time_series=True)
# Conduct training and evaluation inside the trainer
>>> log = trainer.train()
\end{lstlisting}

\subsubsection{Extensions}

Lastly, our package can be easily extended to incorporate multi-task and transfer learning algorithms. We describe two examples below.

For multitask learning, we consider two task grouping methods, which identify related tasks likely to benefit from joint training and optimize them within a shared neural network. These methods include task affinity grouping (TAG) and approximating higher-order task groupings (HOA). Our package can be extended to incorporate these methods with a few lines of code. As shown in Code Snippet \ref{code_multitask_grouping}, users can generate task groupings using the selected method and then apply the multitask trainer to train a model for each group of tasks.

\begin{lstlisting}[language=Python, caption=Training multitask learning models on groups of tasks, label=code_multitask_grouping]
>>> from ml_for_road_safety import MultitaskTrainer
# Generate task groupings from previous task grouping methods, such as TAG or HOA
>>> task_list = ["DE_accident_classification", "IL_accident_classification", "MA_accident_classification", "MD_accident_classification", ...]
>>> task_groups = group_tasks(method = "hoa", task_list)
# Generated task grouping is a list of grouped tasks
>>> task_groups = [
        ["DE_accident_classification", "IL_accident_classification", ...],
        ["MA_accident_classification", "MD_accident_classification", ...],
        ["IA_accident_classification", "NV_accident_classification", ...]
    ]
>>> for group_task_list in task_groups:
>>>     task_datasets = {}; task_evaluators = {}
>>>     for task_name in group_task_list: 
>>>         state_name, data_type, task_type = task_name.split("_")
>>>         task_datasets[task_name] =  TrafficAccidentDataset(state_name)
>>>         task_evaluators[task_name] = Evaluator(task_type)
#       Initialize a trainer with the combined datasets of a group
>>>     trainer = MultitaskTrainer(model, tasks = group_task_list, 
        task_to_datasets=task_datasets, task_to_evaluators=task_evaluators, ...) 
#       Conduct multitask learning on one group of tasks
>>>     trainer.train()
\end{lstlisting}

For transfer learning, we consider two regularization methods for fine-tuning a model trained on a source task to a target task. These methods include soft penalty and sharpness-aware minimization.
Soft penalty regularizes the fine-tuned model distances to the initial model weights. 
Sharpness-aware minimization regularizes the loss sharpness with respect to the model weights.

For example, as shown in Code Snippet \ref{code_advanced_transfer_learning}, a trainer incorporating soft penalty regularization can be implemented by modifying the training loss in the base trainer class. The trainer can then be used to fine-tune a model while penalizing large deviations from the pretrained parameters. Empirically, these techniques improve test performance by an average of $0.6$\% across four states compared to standard fine-tuning. A more comprehensive evaluation of related methods is left for future work.

\begin{lstlisting}[language=Python, caption=Implementation of fine-tuning with soft penalties, label=code_advanced_transfer_learning]
>>> from ml_for_road_safety import Trainer
# Define a trainer for soft penalty inherited from the base Trainer class
>>> class SoftPenaltyTrainer(Trainer):
# Modify the training loss in the training logic
>>>     def train_epoch(self):
#           Combine the soft penalty loss with the cross-entropy loss
>>>         ...
>>>         loss = cross_entropy_loss + \
                   lambda*add_soft_penalty(model, initial_state_dict)
>>>         ...
# Initialize a soft penalty trainer 
>>> trainer = SoftPenaltyTrainer(model, dataset, evaluator, initial_state_dict, ...)
# Conduct training and evaluation inside the trainer
>>> log = trainer.train()
\end{lstlisting}

\section{Additional Related Work Discussions}

\textbf{Spatiotemporal graph neural networks.}
Previous works have proposed spatiotemporal graph neural networks to capture spatial and temporal dependencies for time series analysis on graph-structured data.
DCRNN \cite{li2017diffusion} captures the spatial
dependency using bidirectional random walks on the graph, and the temporal dependency using the encoder-decoder architecture with scheduled sampling.
Instead of applying regular convolutional and recurrent units, STGCN \cite{yu2017spatio} builds the model with complete convolutional structures, which enable much faster training speed with fewer parameters. Another focus is on automatically learning node connections from data. For example, AGCRN \cite{bai2020adaptive} designs two adaptive modules for traffic
forecasting on top of GCN, including one module to capture node-specific patterns and another to infer the inter-dependencies among different traffic series automatically.
Graph WaveNet \cite{wu2019graph} develops an adaptive dependency matrix to capture the hidden spatial dependency in the data and a dilated 1D convolution component to handle very long sequences. 
We refer interested readers to a comprehensive survey \cite{wang2020deep} on spatio-temporal models.

\paragraph{Graph contrastive learning.}
The use of contrastive learning on graph-structured data has been extensively explored in recent work \cite{you2020graph}.
A refined optimization algorithm is introduced to automatically select data augmentations on specific graph data \cite{you2021graph}.
In the domain of spatiotemporal graph learning, contrastive learning has been applied to predict fine-grained urban flows, by designing contrastive losses in both spatial and temporal dimensions \cite{qu2022forecasting}.
AutoST \cite{zhang2023automated} designs a contrastive learning approach for learning spatio-temporal graphs, with a new heterogeneous graph neural network architecture and spatio-temporal augmentation methods. 

\paragraph{Explainability in graph neural networks.}
A recent survey \cite{yuan2022explainability} provides unified and taxonomic explanations regarding the importance of a node/an edge/a subgraph in a graph neural network, etc.
In our leave-one-out analysis, because we are interested in which categories of information (graph structural vs. weather vs. traffic volume) are most useful, our analysis can be viewed as a first-order explanation of the importance of graph structural features.
Further applying the GNN explanation methods \cite{yuan2022explainability} to explain our findings is a research question for future studies.

\paragraph{Multitask and transfer learning.}
Our approach of applying multitask learning is based on recent developments regarding the modeling of negative transfer in multitask learning \cite{yang2020analysis}.
For example, a surrogate modeling approach has been proposed to approximate multitask predictions on various combinations of tasks \cite{li2023identification}.
Building on this approach, a boosting procedure has been designed to construct an ensemble of models for multitask learning on graph-structured data \cite{li2023b}.
Our methods for transfer learning are based on recent developments for robust fine-tuning \cite{li2021improved,ju2022robust}.
In particular, recent work \cite{ju2023generalization} develops a spectrally-normalized generalization bound for graph neural networks and designs a noise stability optimization algorithm for improved fine-tuning.

\paragraph{Dataset development.}
Our dataset includes the road network features and weather information for all the states in the US. The bottleneck is the traffic volume and the accident records. For the eight states in our current dataset, both types of data are published by the Department of Transportation on the respective state’s website. See the links to each state’s government website in Table \ref{table_data_source}.

For the other states, we have checked their Department of Transportation websites, and we could not find detailed data, including accidents and traffic volume (like the eight states we currently have). Once the data is updated, we would be happy to update our dataset as well.

For a few states, for example, California and New York, the traffic volume data and accident information are both available for a few counties through their transportation departments, such as Los Angeles and New York City.
For New York City, we have collected 2.02 million accident records from 2012 to 2023, including the latitude and longitude of each accident.
For California, we have 0.4 million Motor Vehicle Crashes from 2016 to 2021. However, these do not have the latitude and longitude information, so we cannot match a record to a particular edge of the network.

\end{document}